\documentclass[%
 reprint,
superscriptaddress,
 aps,
]{revtex4-2}

\usepackage{hyperref}
\usepackage{multirow}       
\usepackage[caption=false]{subfig}
\usepackage{graphicx}
\usepackage{amsmath}        
\usepackage{amssymb}        
\usepackage{physics}        
\usepackage{xcolor}

\newcommand{\fullpartial}[2]{\frac{\partial #1}{\partial #2}} 
\newcommand{\oprod}{\mathop{\overleftarrow{\prod}}} 
\newcommand{\dif}[1]{\mathop{}\!\mathrm{d}#1~}      
\newcommand{\adj}[1]{#1^\dagger}    

\newcommand{\ha}{\mathrm{h.c.}}     
\newcommand{\an}{\hat a}            
\newcommand{\at}{\adj{\hat a}}      
\newcommand{\ns}{\mathrm{~ns}}       
\newcommand{\Ha}{\mathrm{~Ha}}       
\newcommand{\MHz}{\mathrm{~MHz}}     
\newcommand{\GHz}{\mathrm{~GHz}}     
\newcommand{\REF}{\mathrm{0}}     
\newcommand{\x}{\vb*{\theta}}       
\newcommand{\Ds}{s}                 
\newcommand{\Dsmin}{\Ds_{\min}}     
\newcommand{\eMET}{T_0}             
\newcommand{\EvT}{``error-vs-time''}  
\newcommand{\CvT}{``iterations-vs-time''}   
\newcommand{\finerlabel}{\qty{\alpha\beta}_2}
\newcommand{\finestlabel}{\qty{\alpha\beta}_\infty}

\usepackage{soul}   

\newcommand{\VTCHEMISTRY}{Department of Chemistry, Virginia Tech, Blacksburg, VA 24061}
\newcommand{\VTPHYSICS}{Department of Physics, Virginia Tech, Blacksburg, VA 24061}
\newcommand{\VTQUANTUM}{Virginia Tech Center for Quantum Information Science and Engineering, Blacksburg, VA 24061, USA}

\begin{document}

\preprint{APS/123-QED}


\title{Parameterization and optimizability of pulse-level VQEs} 

\author{Kyle M. Sherbert}
\affiliation{\VTCHEMISTRY}
\affiliation{\VTPHYSICS}
\affiliation{\VTQUANTUM}
\author{Hisham Amer}
\author{Sophia E. Economou}
\author{Edwin Barnes}
\affiliation{\VTPHYSICS}
\affiliation{\VTQUANTUM}
\author{Nicholas J. Mayhall}
\email{nmayhall@vt.edu}
\affiliation{\VTCHEMISTRY}
\affiliation{\VTQUANTUM}


\begin{abstract}
    In conventional variational quantum eigensolvers (VQEs), trial states are prepared by applying  series of parameterized gates to a reference state,
        with the gate parameters being varied to minimize the energy of the target system.
    Recognizing that the gates are  
        intermediates
        which are ultimately compiled into a set of control pulses to be applied to each qubit in the lab,
    the recently proposed ctrl-VQE algorithm takes
        the amplitudes, frequencies, and phases of the pulse
         as the variational parameters used to minimize the molecular energy.
        In this work, we explore how all three degrees of freedom interrelate with one another.
    To this end, we consider several distinct strategies to parameterize the control pulses,
        assessing each one through numerical simulations of a transmon-like device.
    For each parameterization, we contrast the pulse duration required to prepare a good ansatz,
        and the difficulty to optimize that ansatz from a well-defined initial state.
    We deduce several guiding heuristics to implement practical ctrl-VQE in hardware,
        which we anticipate will generalize for generic device architectures.
\end{abstract}

\maketitle


\section{Introduction}
Variational quantum eigensolvers (VQEs)
    are among the most promising candidates for achieving useful computations
    in chemistry on near-term quantum computers~\cite{bauer_quantum_2020, mcardle2020quantum, cao2019quantum, fedorov2022vqe, tilly_variational_2022, cerezo2021variational}.
At their core, they are predict-and-test methods,
    where a quantum state, determined by a set of classical parameters as specified by an {\em ansatz},
    is prepared on the quantum computer, and its energy measured.
Then a new quantum state is prepared with a new set of parameters,
    selected by any classical optimization algorithm,
    and the process is repeated until the energy is minimized.
This minimal energy is an upper-bound to the ground-state energy of the system,
    and, if the optimization is carried out perfectly,
    the final state is the best approximation to the ground state
    attainable with the chosen ansatz.
    
Clearly, a key decision in any VQE is the choice of ansatz,
    typically expressed as a series of parameterized unitary {\em gates} in a quantum circuit.
While the original formulation of VQE utilized an ansatz inspired by the underlying chemistry~\cite{peruzzo2014variational},
    many novel ans\"atze have been proposed which balance chemical intuition
    with information about which parametric and entangling operations are most easily implemented on the device~\cite{kandala2017hardware, lee2018generalized, ryabinkin2018qubit, gard2020efficient, grimsley2019adaptive, yordanov_qubit_2021, tang2021qubit}.
In either case,
    for experimental realization, the quantum circuit must be compiled into a sequence of electromagnetic pulses that are applied to the device. For transmon-type devices (those considered in this work), the control pulses are
    microwave signals which drive the states of each physical qubit
    as prescribed by each unitary gate in the circuit.
Unfortunately, if each gate is to be implemented with high fidelity in modern hardware,
    even the most compact ans\"atze for moderately-sized systems
    typically compile into control pulses with a duration far exceeding the coherence time of the device~\cite{Meitei2020}.

Three of the authors previously proposed the algorithm ctrl-VQE~\cite{Meitei2020,asthana_leakage_2023}, which
    takes the idea of a hardware-efficient ansatz to the extreme
    by parameterizing the actual physical control pulses used in the lab,
    bypassing the use of gates entirely.
Designing the ansatz at the pulse level allows drastically shorter evolution times,
    even approaching the quantum speed limits imposed by the hardware \cite{deffner2017quantum},
    and hypothetically enabling the VQE to study much larger or more complex systems.
The methods and mathematical theory behind ctrl-VQE
    are built upon the closely related field of quantum control,
    and we refer the reader to Refs.~\cite{liberzon2011calculus,Ansel_2024}
    for  introductions to this topic.
Recently, several works have  applied pulse-level design to estimate eigenvalues of a target operator, in line with the original ctrl-VQE~\cite{magann2021pulses, leng_differentiable_2022, meirom_pansatz_2023, liang_towards_2023, egger_pulse_2023, kottmann_evaluating_2023, keijzer_pulse_2023, entin_molecular_2024,long2024minimalevolutiontimesfast}, including a handful with proof-of-concept hardware demonstrations on simple systems.

Most of these efforts have focused on parametric modifications of established one- and two-qubit gates,
    while our original work has emphasized \textit{de novo} pulses which act more like a multi-qubit gate.
    In fact, ctrl-VQE represents the most flexible variational ansatz possible for any given hardware. 
While this approach is very powerful,
    it presents the challenge that there is an overwhelming number of accessible degrees of freedom.
There is typically one or more control pulse simultaneously applied to each qubit,
    and each control pulse is parameterized by an amplitude ($A$), a phase ($\phi$), and an (angular) frequency ($\nu$), each of which may be varied in time,
    resulting in an ostensibly infinite number of parameters (before hardware bandwidth limits are considered). An important open question is which parameters are most important for quickly preparing a target state while avoiding optimization issues.
    
In this paper, we investigate the impact that each of the pulse parameters tends to have and establish useful heuristics for practical pulse parameterizations.
For example, we present empirical evidence that representing the amplitude $A$ and phase $\phi$ together as a complex amplitude with real and imaginary components enables notably more efficient optimizations than when varying each pulse parameter independently.
Furthermore, we find that varying the phase $\phi$ is likely to be far more effective than varying drive frequency $\nu$ in a practical ctrl-VQE experiment,
    since varying the latter tends to significantly impede the efficiency of optimization, without providing any commensurate improvement in controllability or minimal evolution time.
We mainly focus on transmon-type architectures in this work, but we anticipate our conclusions can guide pulse-level VQE experiments on a wide array of devices.

The paper is organized as follows:
In Section~\ref{sec:theory}, we review the mathematical basis of the ctrl-VQE algorithm, and we introduce the various choices of parameterization for our ansatz.
In Section~\ref{sec:methods}, we briefly describe the methods and parameters used to generate the data presented in this paper.
In Section~\ref{sec:results}, we present a series of experiments contrasting the optimizability of each parameterization.
Section~\ref{sec:conclusions} summarizes our results and condenses our findings into a set of guidelines for experimental realizations of ctrl-VQE.

\section{Theory}
\label{sec:theory}

We take as our starting point a quantum observable $\hat O$ (e.g., a molecular Hamiltonian),
    and a reference state $\ket{\psi_{\REF}}$ (e.g., the Hartree-Fock state),
    both mapped onto qubits via a suitable transformation (e.g., Jordan-Wigner).
Our variational ansatz $\ket{\psi(\x)}$ will take the form:
\begin{equation}
    \ket{\psi(\x)} = U(\x) \ket{\psi_{\REF}},
\end{equation}
where $U(\x)$ is the unitary state-preparation operator.
The expectation value of $\hat O$ is the energy $E(\x)$:
\begin{equation}
    E(\x) \equiv \expval{\hat O}{\psi(\x)},
\end{equation}
which is the function that the optimization routine in ctrl-VQE will attempt to minimize.

In ctrl-VQE, the unitary $U(\x)$ is the physical evolution under a device Hamiltonian
    (as distinct from traditional VQEs, where it is typically a sequence of logical operations on the qubit space).
The device Hamiltonian may be separated into a static component $\hat H_0$ and a drive component $\hat V(t)$.
Variational parameters are typically associated with the time-dependent drive $\hat V(t)$.
Frequently, $\hat V(t)$ is comprised of several independent drives;
    in this work, we consider an independent drive $\hat V_q(t)$ applied to each qubit $q$.
As such, our choice of ansatz in ctrl-VQE is essentially the unitary time evolution operator:
\begin{equation}
    U(\x) = \mathcal{\hat T} \exp(-i \int_0^T \dif{t} \qty[\hat H_0 + \sum_q \hat V_q(\x, t)]),
\end{equation}
where $\mathcal{T}$ is the time-ordering operator and $T$ is the total pulse duration.
While the underlying physical mechanism varies for different quantum computing architectures, the drive term in a qubit Hamiltonian can typically be modeled as an interaction between a dipole moment (the qubit) and a classical electric field (the {\em control field}):
\begin{equation}
    \label{eqn:realdrive}
    \hat V_q(t) \propto D(t) (\an_q+\at_q),
\end{equation}
where $\an_q$ is the bosonic lowering operator acting on the transmon qubit state.
The electric field arises from an applied voltage that
    oscillates sinusoidally at microwave frequencies:
\begin{equation}
    \label{eqn:EM}
    D(t) = A \cos(\nu t + \phi).
\end{equation}
The parameters we have control over are therefore the amplitude $A$,
    the frequency $\nu$,
    and the phase $\phi$ of the field,
    each of which may themselves be functions of time.

When the amplitude is relatively weak with respect to the drive frequency,
    and when the difference between the drive frequency and the qubit frequency is negligible compared to either of these quantities,
    the rotating wave approximation can be used to convert the sinusoidal form of Eq.~\eqref{eqn:EM} into a complex phase:
\begin{equation}
    \label{eqn:drive}
    \hat V_q(t) \xrightarrow{\mathrm{RWA}} \Omega e^{i\nu t} \an_q + \ha ,
\end{equation}
where the complex amplitude $\Omega$ is defined in polar notation by the amplitude (i.e., modulus) and phase:
\begin{equation}
    \label{eq::polar}
    \Omega \equiv A e^{i\phi} /2.
\end{equation}
Alternatively, we could parameterize our control field equivalently in a Cartesian notation, defining the real-valued amplitudes $\alpha$ and $\beta$ as the real and imaginary parts of $\Omega$:
\begin{equation}
    \label{eq::cartesian}
    \Omega \equiv \alpha  + i \beta.
\end{equation}
This Cartesian parameterization is often more convenient in software implementations.

In the absence of any control field, a single qubit will evolve under its static Hamiltonian $H_0$, which in transmons can be modeled as an anharmonic oscillator
    with a characteristic frequency $\omega$
    describing the energy gap between the lowest two states $\ket{0}$ and $\ket{1}$.
In such cases, it is common to make a coordinate transformation from the lab frame
    into the rotating frame of the device.
\begin{equation}
    U^{(t)}_{\rm RF \leftarrow LAB} = e^{i H_0 t}
\end{equation}
The Hamiltonian in this frame is written as:
\begin{equation}
    \label{eqn:rotatingdrive}
    V_R(t) = \Omega e^{i\Delta t} \an + \ha ,
\end{equation}
where the detuning $\Delta$ is simply the difference
    between the drive frequency $\nu$ and the qubit frequency $\omega$:
\begin{equation}
    \Delta \equiv \nu - \omega,
\end{equation}
so that $\Delta=0$ corresponds to driving on resonance.
Note the similarity of Eq.~\ref{eqn:rotatingdrive} to Eq.~\ref{eqn:drive} when restricted to a single qubit.
The total Hamiltonian in the rotating frame is equivalent to the drive term of the lab frame, after replacing the drive frequency $\nu$ with the detuning $\Delta$.
In multi-qubit systems, static coupling between qubits leads to a dressing of the bare qubit parameters,
    but the detuning $\Delta$ of any given pulse from the frequency $\omega$ of the qubit it targets
    remains a useful parameter.
Although the drive frequency $\nu_q$ for each qubit is the physical parameter most directly changed in the lab,
    the detuning $\Delta_q$ is equivalent up to a constant and is the more intuitive parameter to work with in the rotating frame, so we adopt $\Delta$ as a variational parameter for the remainder of this work.

In our previous work~\cite{Meitei2020,asthana_leakage_2023},
    we took the amplitude $\Omega$ to be real,
    essentially restricting the phase $\phi$ to either 0 (when $\Omega\ge0$) or $\pi$ (when $\Omega<0$). In the present work, we instead consider various parameter sets, including complex amplitudes in some cases; all parameter sets are
    listed in Table~\ref{tab::parameterizations}.
Each parameter can be varied in an optimization loop, between control pulses.
Furthermore, each parameter could in principle 
    be varied as an arbitrary function in time, throughout the pulse duration.
As a practical matter,
the time scale at which the drive frequencies can be varied in hardware is somewhat larger than the one for amplitudes and phases, and the frequencies are typically restricted to discrete values during the course of a single pulse schedule, so
    we will take $\Delta$ as time-independent for any given control pulse, as in our previous papers.
For simplicity, we will limit ourselves to ``windowed'' pulses
    where each window has a constant amplitude and phase;
    arbitrary pulse shapes can be approximated
    by taking windows of arbitrarily short duration.

\begin{table}
    \begin{tabular}{|c|c|c|c|}
        \hline
        Label & \centering $\Omega$ & \centering $\nu$ & Figs \\ \hline
        $\qty{\alpha\beta}$ & 2 param. per window, Cartesian & Resonant. & \ref{fig::LiH}, \ref{fig::windowspacing}, \ref{fig::polarcartesian}, \ref{fig::resonantdetuned} \\
        \hline
        $\finerlabel$ & \multicolumn{2}{c|}{Same as $\qty{\alpha\beta}$, twice as many windows.} & \ref{fig::windowspacing} \\
        $\finestlabel$ & \multicolumn{2}{c|}{Same as $\qty{\alpha\beta}$, continuous windows.} & \ref{fig::windowspacing} \\
        \hline
        $\qty{A\phi}$ & 2/window, polar & Resonant. & \ref{fig::polarcartesian} \\
        \hline
        $\qty{\alpha}$ & 1/window, real only & Resonant. & \ref{fig::resonantdetuned} \\
        $\qty{\alpha\Delta}$ & 1/window, real only & 1/pulse & \ref{fig::resonantdetuned} \\
        $\qty{\alpha\beta\Delta}$ & 2/window, Cartesian & 1/pulse & \ref{fig::resonantdetuned} \\
        \hline
    \end{tabular}
    \caption{Summary of pulse parameterizations considered in this paper. Except when otherwise noted, optimizations begin with all complex amplitudes $\Omega$ initialized to zero, and drive frequencies $\nu$ initialized on resonance (i.e., $\Delta=0)$.}
    \label{tab::parameterizations}
\end{table}

\section{Methods}
\label{sec:methods}

The results presented in this paper focus on the lithium hydride molecule (LiH) mapped onto four qubits as a case study with a nuclear separation of 3.0\AA.
This is a stretched geometry where the mean-field solution has a 72\% overlap with the ground state (as compared to the equilibrium geometry of 1.5\AA, where the mean-field solution has a 98\% overlap with the ground state).
This system is small but strongly correlated, making it a useful case study for probing the efficiency of pulse-level VQE.

We use the \texttt{pyscf} package
    \cite{pyscf}
    to calculate electronic integrals for each molecular geometry,
    and we use the \texttt{qiskit} package
    \cite{qiskit}
    to construct the second-quantized molecular Hamiltonian,
    map it onto qubits with the parity mapping, and apply two-qubit reduction.
    \cite{bravyi_tapering_2017}
We use the Hartree-Fock singlet state as our reference,
    and we represent the Hamiltonian as a matrix in the Hartree-Fock basis
    so that the reference state is a computational basis state.

We use our own Julia code
    \cite{julia,CtrlVQE}
    to simulate unitary time evolution
    under the rotating wave approximation with arbitrary control pulses,
    using Trotterized time evolution with twenty time steps per nanosecond.
Before the pulse, the qubits are prepared into the Hartree-Fock reference state,
    and after the pulse, the resulting statevector (in the rotating frame)
    is used to measure the energy, i.e., the expectation value of our molecular Hamiltonian.
We measure in the rotating frame so that when all pulse amplitudes are set to zero, the cost function (molecular energy) does not change in time.

The static Hamiltonian $H_0$ is modeled as a set of coupled harmonic oscillators,
    with each qubit truncated to the first two energy levels:
\begin{equation}
    \hat H_0 = \sum_q \omega_q \at_q \an_q + \sum_{\expval{p,q}} g_{pq} (\at_p \an_q + \at_q \an_p).
\end{equation}
Note that this is a common model for transmon devices,
    \cite{Krantz_2019,Roth_2023}
    excepting an additional anharmonicity term $\at_q \at_q \an_q \an_q$
    which is dropped in the two-level approximation.
In this paper, coupling strengths are fixed to $g_{pq}=0.02 \cdot 2\pi\GHz$ if $q=p+1$ (i.e., linear nearest-neighbor coupling),
    and qubit frequencies are equally spaced such that $\omega_q=(4.80 + 0.02q) \cdot 2\pi\GHz$.
These parameters roughly match those found in real IBMQ devices,
    while being systematically scalable to larger systems.

We use the \texttt{Optim.jl} implementation of BFGS (a quasi-Newton optimization algorithm)
    \cite{optim}
    with analytical gradients
    (see Appendix~\ref{app::gradient} and \cite{khaneja_optimal_2005})
    to iteratively minimize the molecular energy with respect to a specified drive parameterization.
Hardware bounds on the maximum amplitude are enforced
    by including a smooth penalty term in the cost-function on any time step
    for which the pulse amplitude exceeds $\Omega_{\rm max}=0.02 \cdot 2\pi\GHz$.
Optimization terminates when the maximum of the analytical gradient
    has converged to within 1e-6.

Our methodology adopts the following approximations for the sake of computational efficiency: (1) the Rotating Wave Approximation, (2) Trotterization, and (3) two-level truncation.
Benchmarks indicate that, in the regimes considered here, the rotating wave approximation and Trotterization each incur errors on the order of $1e-4\Ha$.
Further, pulses optimized in the two-level approximation result in significant leakage when simulated with more levels (convergence requires as many as five levels in these regimes).
Finally, the device parameters we simulate do not correspond exactly to any existing device.
Therefore, we emphasize that the specific pulses and optimization trajectories we report in this paper should not be taken as absolutes.
Rather, we will emphasize the {\em relative differences} in pulse and optimization trajectories when comparing different parameterizations, differences which we expect would also be observed on similar hardware.

\section{Results}
\label{sec:results}

Previous works on ctrl-VQE have emphasized the role of the amplitude $A$
    as a function of time.
In this work, we will assess the relative significance
    of the phase $\phi$ and the detuning $\Delta$ (capturing the frequency degree of freedom),
    as measured by the pulse duration necessary
    to achieve high-accuracy estimates of a ground state energy.
We will make extensive use of \EvT{} plots 
    and \CvT{} plots, 
    which show how the accuracy and iteration count of a ctrl-VQE optimization change
    as a function of pulse duration.
    
\begin{figure*}
    \subfloat[\label{fig::LiH:energy}]{%
        \includegraphics[width=\columnwidth]{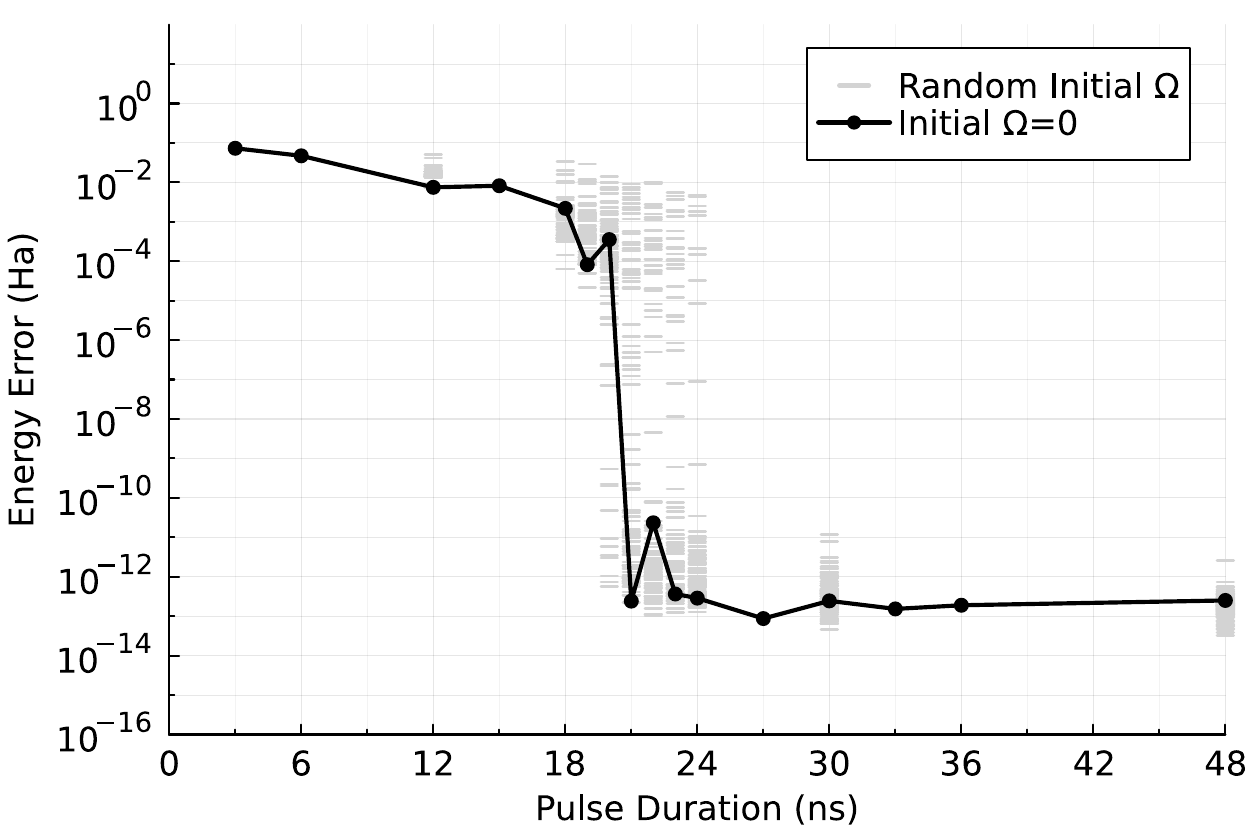}%
    }
    \hfill
    \subfloat[\label{fig::LiH:iterations}]{%
        \includegraphics[width=\columnwidth]{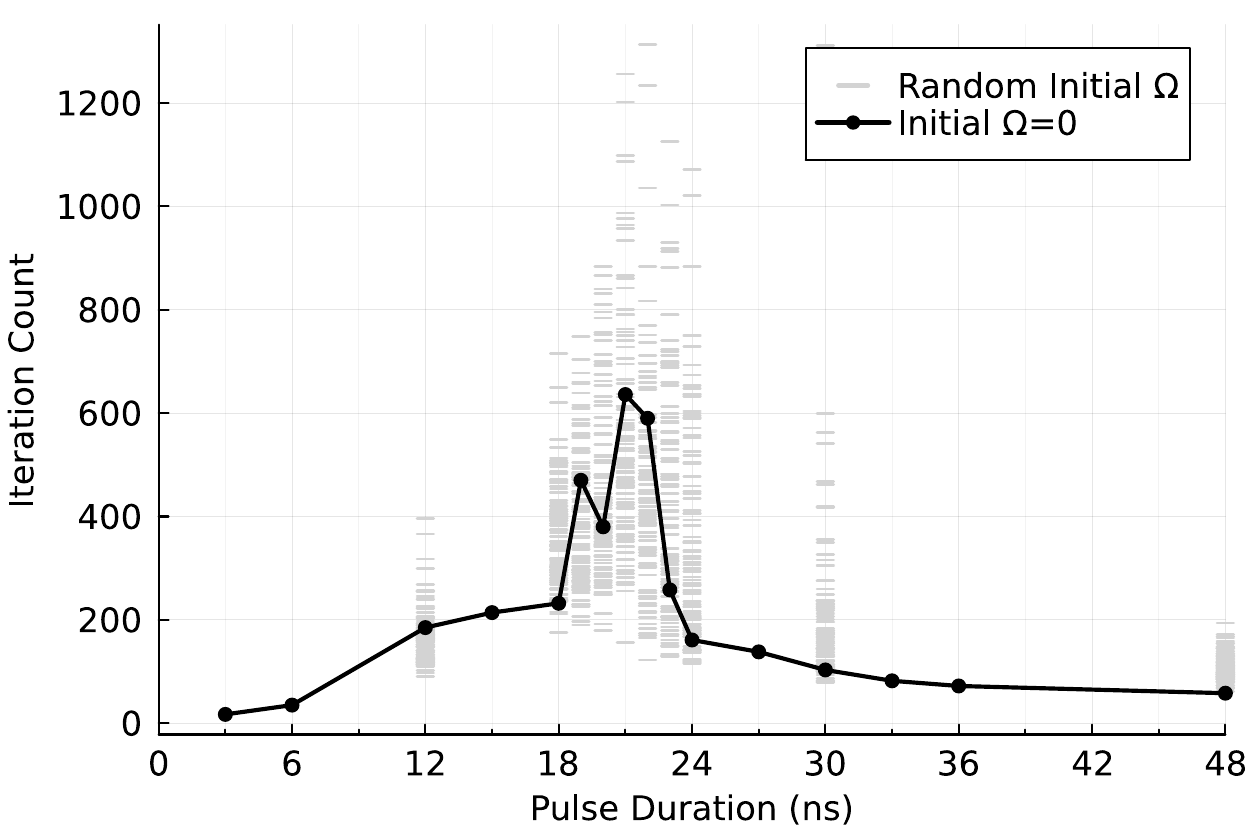}%
    }

    \subfloat[\label{fig::LiH:moduli}]{%
        \includegraphics[width=\columnwidth]{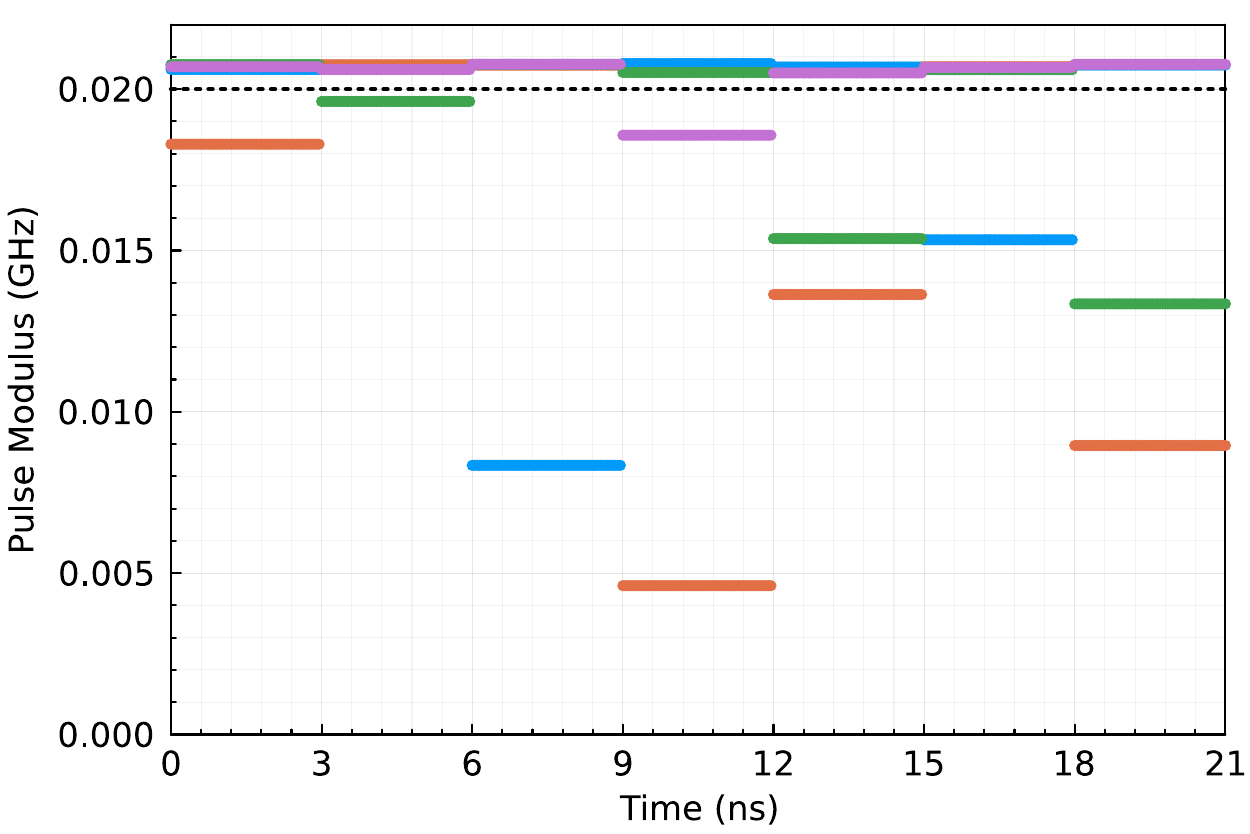}%
    }
    \hfill
    \subfloat[\label{fig::LiH:phases}]{%
        \includegraphics[width=\columnwidth]{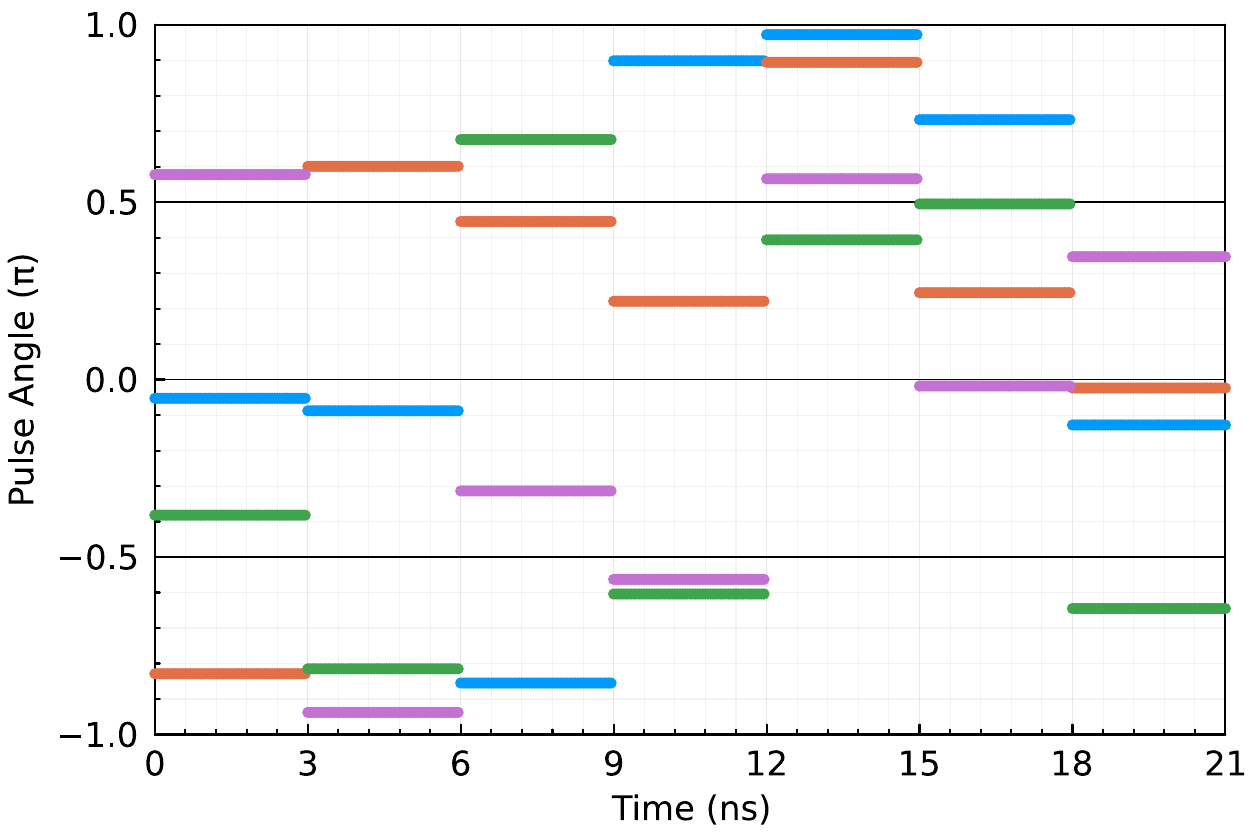}%
    }
    
    \caption{
        (\ref{fig::LiH:energy}) Ground state energy error vs. pulse duration for a ctrl-VQE simulation of LiH with nuclear separation 3.0\AA, using the $\qty{\alpha\beta}$ parameterization. Each pulse is divided into as many uniformly spaced windows as possible such that window length satisfies $\Ds\ge3.0\ns$. For the black curve, optimization is carried out with all parameters starting at zero. Gray dashes mark optimized runs when parameters are initialized with random values.
        (\ref{fig::LiH:iterations}) The number of BFGS iterations needed to obtain the energies in (\ref{fig::LiH:energy}).
        (\ref{fig::LiH:moduli}) Optimized pulse amplitudes $A(t)$ for the ctrl-VQE run at $T=21\ns$. Each color corresponds to the drive on a different qubit. The dashed line marks the bound above which penalties are imposed, ostensibly the maximum drive allowed by the device.
        (\ref{fig::LiH:phases}) Optimized pulse phases $\phi(t)$ for the ctrl-VQE run at $T=21\ns$. The solid lines occur at intervals of $\pi/2$, separating quadrants of the unit circle.
    }
    \label{fig::LiH}
\end{figure*}

Each data point in these plots is obtained by simulating ctrl-VQE
     to optimize parameters in a pulse with fixed duration $T$.
At the start of optimization, pulse parameters are initialized to zero,
    meaning the pulse shape starts flat and the drive frequencies start on resonance.
The pulse is divided into windows with uniform spacing $\Ds$,
    with longer pulses divided into more windows (details given in Section~\ref{sec::windowspacing}).
In the \EvT{} plots, we plot the difference between the final optimized energy
    and the value obtained from exact diagonalization of the Hamiltonian matrix.
In the \CvT{} plots, we plot the number of iterations the BFGS optimizer requires to converge.
This number is unique to our choice of optimizer,
    but it allows a heuristic comparison of the expected number of circuit evaluations
    required to obtain the accuracies presented in the corresponding \EvT{} plot.

Typical behavior in an \EvT{} plot is for the error to start off large, as seen in Fig.~\ref{fig::LiH:energy}:
When the pulse duration is very small,
    the time evolution is insufficient to drive the system far from the reference state.
As pulse duration is increased, the optimized energy error tends to decrease
    exponentially (appearing linear on the semi-log plot).
However, at some critical time, there is a sharp transition to highly accurate energies,
    where the optimized energy error is on the order of numerical precision;
    increasing pulse duration beyond this point has negligible impact on the accuracy.
The pulse duration for which this transition occurs 
    is closely related to the {\em minimal evolution time} (MET) from optimal control theory,
    which is the absolute shortest time required to prepare one state from another,
    subject to a given set of pulse constraints~\cite{deffner2017quantum,wilhelm2020introduction}.
According to optimal control theory,
    amplitudes will tend to saturate their maximal bounds $\Omega_{\rm max}$ as the pulse duration approaches the MET.
For real-valued amplitudes (the $\qty{\alpha}$ and $\qty{\alpha\Delta}$ parameterizations),
    the MET is achieved with {\em bang-bang} pulse shapes which are characterized as pulses where the amplitude switches between $\pm\Omega_{\rm max}$,
    as observed in our previous work~\cite{asthana_leakage_2023}
    and in a control-theoretic analysis of the quantum approximate optimization algorithm \cite{yang2017optimizing}.
For complex-valued amplitudes, optimal control theory imposes no \textit{a priori} constraints on the phases $\phi$,
    but the amplitude parameters $A$ will tend toward $\Omega_{\rm max}$.
One can see this effect in Figs.~\ref{fig::LiH:moduli} and~\ref{fig::LiH:phases},
    which show the amplitudes and phases optimized from zero at $T=21\ns$,
    the shortest pulse duration in Fig.~\ref{fig::LiH:energy} which results in a ctrl-VQE run obtaining energy errors less than $10^{-8}\Ha$ when all parameters are initialized to zero.

Converging to the true MET is extremely computationally expensive, as it requires, in principle, an exhaustive search over all parameter space. The approach we have taken previously \cite{asthana_leakage_2023} simply samples over multiple optimizations using randomly initialized pulses. 
For example, the gray ticks we plot in Fig.~\ref{fig::LiH:energy} are the optimized energy errors
    from 100 runs when the initial pulse parameters are uniformly sampled from the allowed range.
Several runs with $T=20\ns$ are successful,
    and it may well be that a perfect global optimization
    could locate successful pulses with even lower duration.
However, we choose to use the more computationally tractable zero-pulse initialization
    as a heuristic for the remainder of this work,
    and we will refer to the evolution time in an \EvT{} plot
    as the ``{\em effective}'' minimal evolution time (eMET) $\eMET$.
    
Interestingly, an \CvT{} plot (Fig.~\ref{fig::LiH:iterations}) exhibits an iteration count that
    is relatively low for most pulse durations,
    {\em except} for a broad and large peak near the eMET $\eMET$.
This is due to the saturation of the amplitude bounds observed above,
    straining the optimizer as it balances reductions to the molecular energy
    and to the penalty term enforcing the amplitude bounds.
Therefore, we anticipate real ctrl-VQE experiments
    will want to target pulse durations sufficiently beyond the eMET,
    to achieve the best trade-off between the number of circuit evaluations
    and decoherence error.

In the following subsections, we present \EvT and \CvT plots contrasting each parameterization from Table~\ref{tab::parameterizations}, using our LiH model.
In Appendix~\ref{sec::longersurvey}, we provide preliminary results showcasing the most effective parameterizations with increasing system size.

\subsection{Characterizing eMET for different window lengths}
\label{sec::windowspacing}

\begin{figure}
    \subfloat[\label{fig::windowspacing:energy}]{
        \includegraphics[width=\columnwidth]{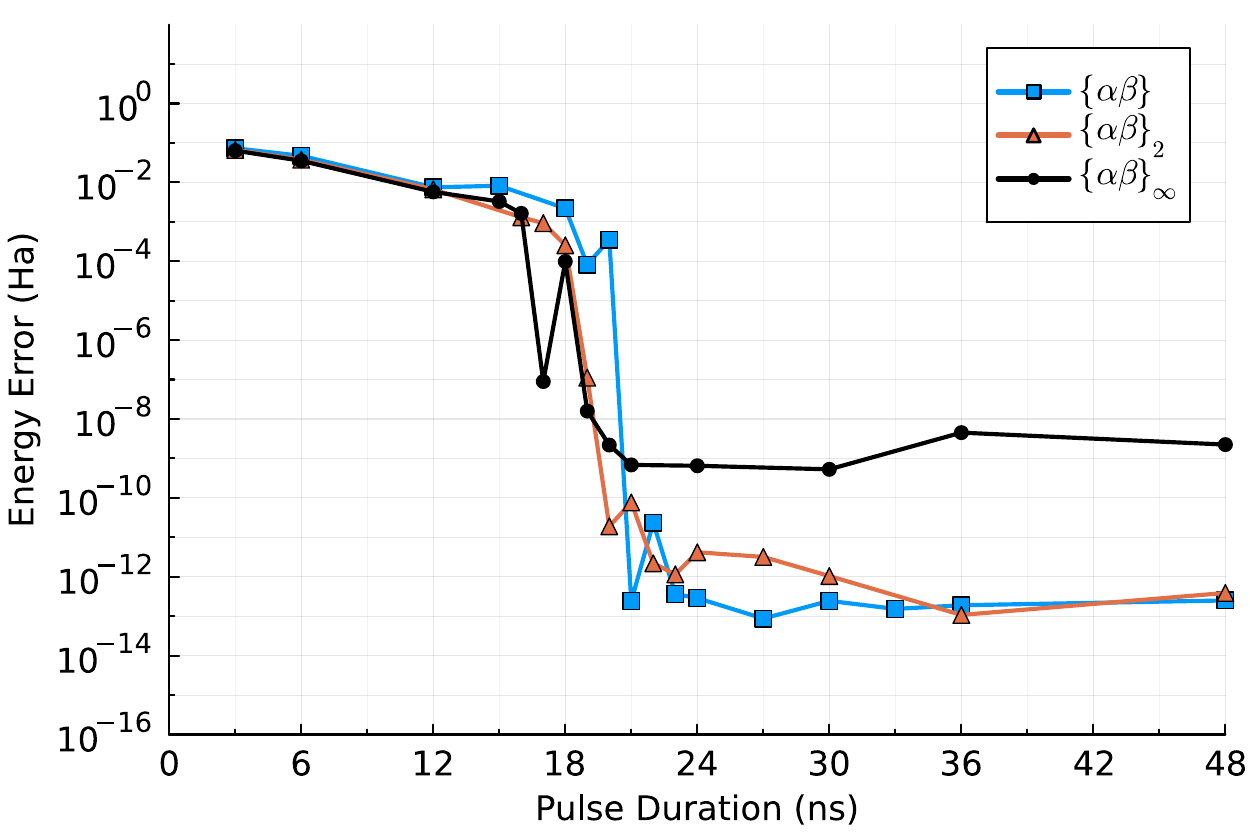}
    }
    \hfill
    \subfloat[\label{fig::windowspacing:iterations}]{
        \includegraphics[width=\columnwidth]{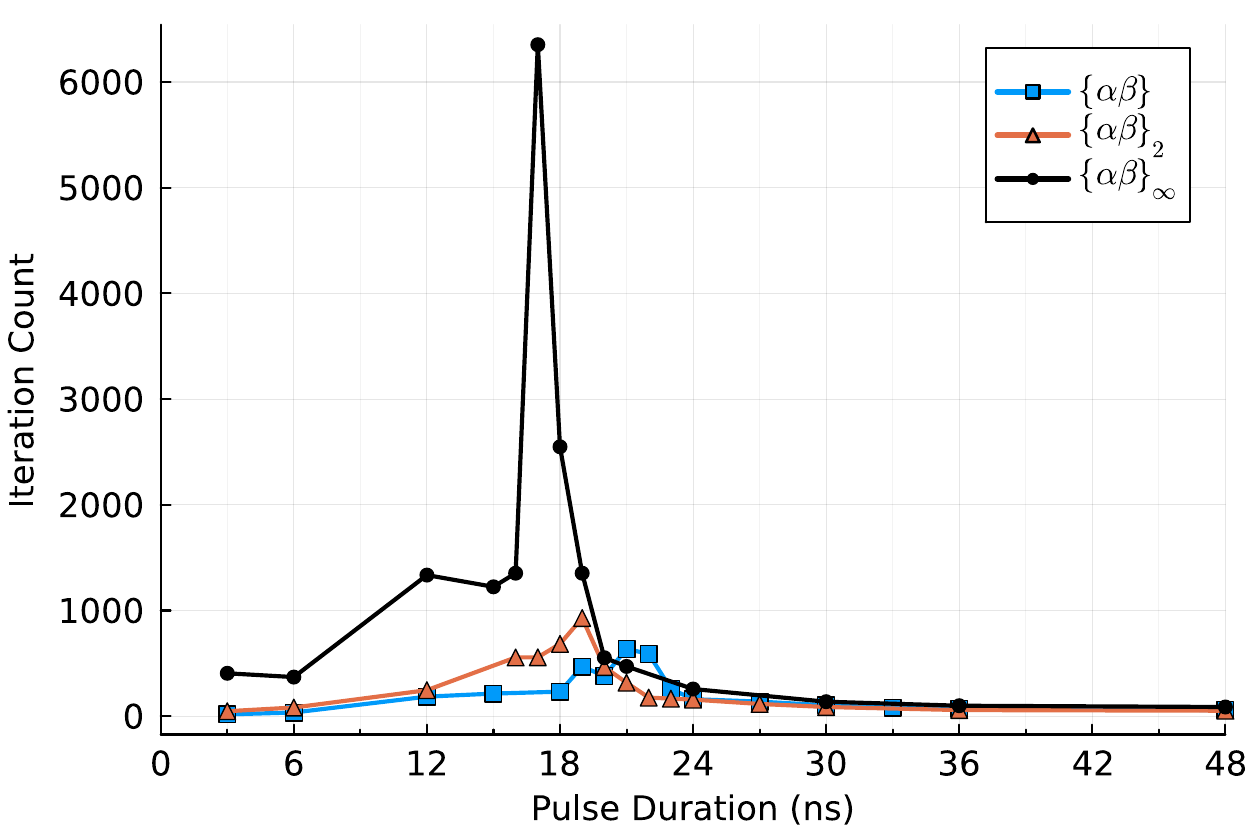}
    }
    \caption{
        (\ref{fig::windowspacing:energy}) Ground state energy error vs. pulse duration of ctrl-VQE applied to LiH with nuclear separation 3.0\AA, using the $\qty{\alpha\beta}$ parameterization with three different numbers of time windows. Each pulse is divided into as many uniformly spaced windows as possible such that the window length $\Ds$ is constrained as labeled. The $\qty{\alpha\beta}_\infty$ simulations use as many windows as there are time-steps in the evolution, such that $\Ds=0.05\ns$. All parameters are initialized to zero in the optimization steps.
        (\ref{fig::windowspacing:iterations}) The number of BFGS iterations needed to obtain the energies in (\ref{fig::windowspacing:energy}).
    }
    \label{fig::windowspacing}
\end{figure}

Throughout this paper, pulses are divided
    into a number of windows chosen such that the duration of each window
    is as close as possible to, but never below, some minimum duration $\Dsmin$.
For example, if $\Dsmin=3\ns$,
    a $T=24\ns$ pulse would be divided into eight windows of $\Ds=3\ns$ each,
    while a $T=23.8\ns$ pulse would be divided into seven windows of $\Ds=3.4\ns$ each.
This choice ensures that longer pulses have more degrees of freedom
    while simultaneously dampening high-frequency components in the control signals
    that are difficult to implement experimentally.
By choosing $\Dsmin=3\ns$, we approximate an effective bandwidth
    of roughly $1/2\Dsmin \sim 167\MHz$ in the microwave pulse generator.

Figure~\ref{fig::windowspacing} shows the \EvT{} and \CvT{} plots
    comparing different choices of $\Dsmin$
    for a LiH molecule with a nuclear separation of 3.0\AA,
    using the parameter set $\qty{\alpha\beta}$
    (complex amplitudes parameterized with Cartesian notation,
    and the drive frequency for each qubit is fixed on resonance).
The ``continuous'' curve $\finestlabel$ assigns independent $\alpha$ and $\beta$ parameters
    for each point in the Trotterized time evolution,
    corresponding to a $\Ds$ on the order of the Trotter step, $\sim0.05\ns$.
Despite the vastly larger number of parameters in this case compared to the other two cases shown in the figure,
    the eMET $\eMET$ changes by only about $3\ns$,
    emphasizing the importance of evolution time over degrees of freedom.
As can be seen in the \CvT{} plot,
    the larger number of parameters does have a significant effect on computational cost,
    so we will adopt the reasonable value of $\Dsmin=3.0\ns$ for most results in this paper.
    
We note that the somewhat higher asymptote observed for energy errors in Fig.~\ref{fig::windowspacing:energy} is an artifact of our optimization routine using the $L_\infty$ norm rather than, say, the $L_2$ norm as the convergence criterion.
Thus, gradient vectors with a very large number of small but non-zero values will be somewhat prematurely flagged as converged.
We anticipate that a correspondingly stricter choice of convergence criterion in the runs with more parameters would result in a taller, broader, and more symmetric \CvT curve than the one observed in Fig.~\ref{fig::windowspacing:iterations}, though we have not confirmed this numerically.

\begin{figure}
    \subfloat[\label{fig::pathlengths:fidelity}]{
        \includegraphics[width=\columnwidth]{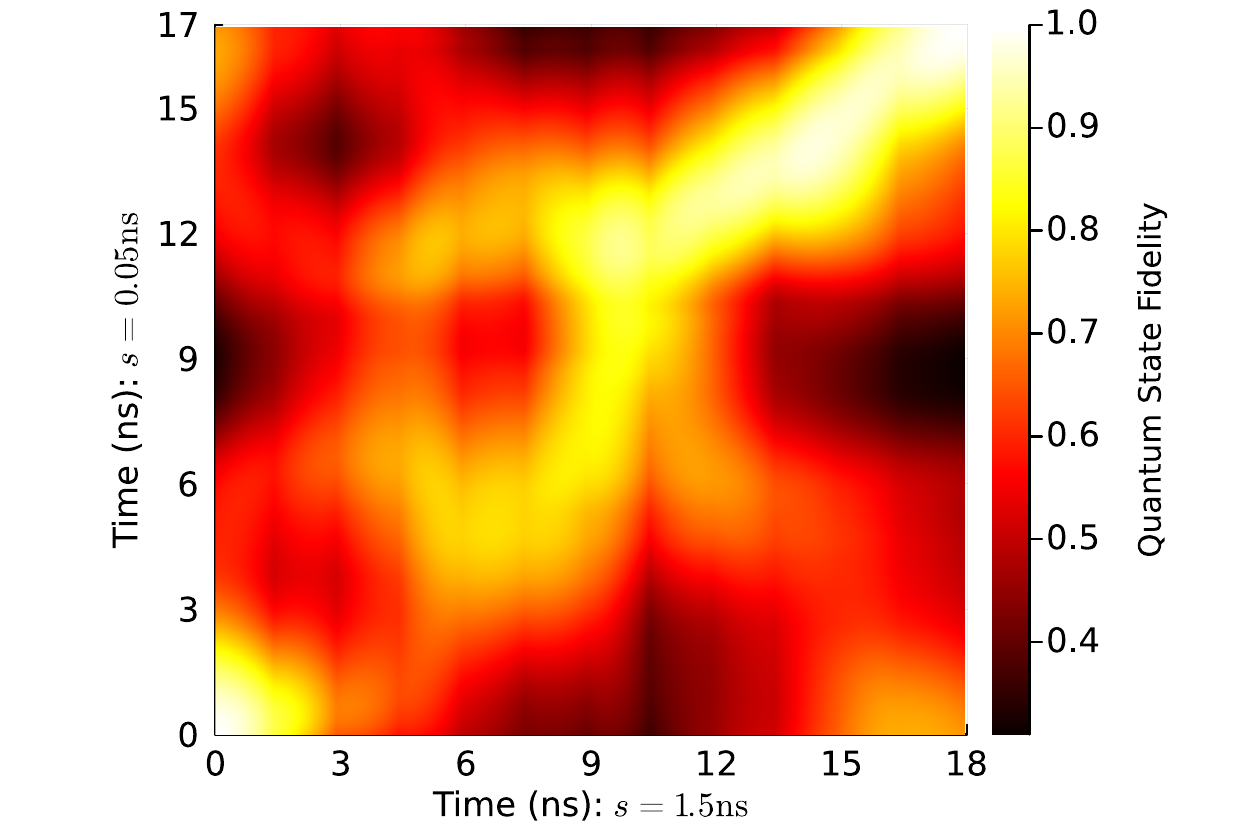}
    }
    \hfill
    \subfloat[\label{fig::pathlengths:cartoon}]{
        \includegraphics[width=0.5\columnwidth]{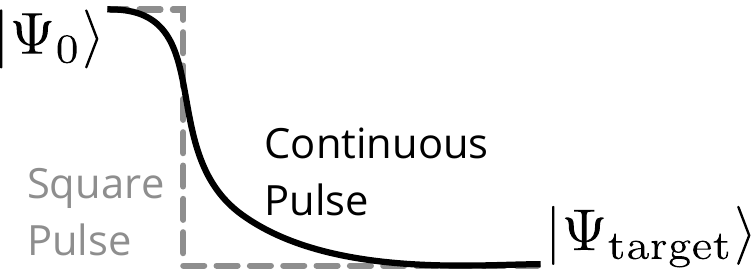}
    }
    \caption{
        (\ref{fig::pathlengths:fidelity}) The fidelity between quantum states as a function of time, following trajectories solving LiH with nuclear separation 3.0\AA, from two optimized pulses: a $17.0\ns$ pulse where parameters were varied quasi-continuously, and an $18.0\ns$ pulse where parameters were restricted to window lengths of $\Ds = 1.5\ns$ each. Bright colors indicate high overlap at the corresponding points in the trajectory, while black indicates orthogonal states.
        (\ref{fig::pathlengths:cartoon}) A cartoon depiction of how finite window lengths affect evolution time. When pulse parameters are allowed to vary continuously in time, denoted by the black curve, ctrl-VQE can in principle find the shortest possible path between the reference state and the ground state. When parameters are constrained to be constant over finite windows, ctrl-VQE can still approximate this ideal path. The total path length, and therefore the total evolution time, will be marginally longer, but the main contribution to evolution time is the length of the ideal path.
    }
    \label{fig::pathlengths}
\end{figure}

We can gain an intuitive understanding for why
    the number of parameters does not have a significant impact on the eMET $\eMET$
    by studying the cartoon in Fig.~\ref{fig::pathlengths:cartoon}:
When the complex amplitude $\Omega$ varies continuously in time, the optimized pulses
    generate the shortest possible path through Hilbert space
    between the reference state and the ground state.
The time it takes to traverse this path is determined by the {\em Quantum Speed Limit},
    which dictates that the rate at which a system can move through Hilbert space
    is bounded by the norm of the drive Hamiltonian,
    determined in our case by the drive amplitudes $A$~\cite{deffner2017quantum}.
Thus, hardware constraints on the maximal drive amplitude $A$ for each pulse
    impose a {\em fundamental} MET $T_{0}$ required to prepare our ground state.
When we impose finite windows on the pulse shape,
    the optimal path traced through Hilbert space is perturbed from the optimum.
Longer windows result in larger perturbations,
    and therefore longer evolution times.
However, as long as the window duration $\Ds$ is relatively small
    with respect to the pulse duration $T$,
    the dominant contribution to the evolution time remains
    the original fundamental MET $T_{0}$.

Fig.~\ref{fig::pathlengths:fidelity} provides a particularly striking visualization of this perturbation.
First, we take two optimized pulses which successfully prepare the target state in the least possible pulse duration $T$, one with $\Ds=0.05\ns$ (``continuous'') achieving $T=17\ns$, and one with $\Ds=1.5\ns$ achieving $T=18\ns$.
We note that optimization of the $\Ds=1.5\ns$ case required randomly initializing pulse parameters to find a successful pulse with $T=18.0\ns$, as opposed to the data presented in Fig.~\ref{fig::windowspacing}, which reports results when initializing all pulses from zero.
Next, we calculate the trajectories $\ket{\psi(t)}$, $\ket{\phi(t)}$ in the rotating frame, when applying each pulse.
Finally, for every pair of times $t$, $t'$, we evaluate the fidelity $|\braket{\psi(t)}{\phi(t')}|^2$.
The white spot in the bottom left of Fig.~\ref{fig::pathlengths:fidelity} indicates that both pulses are applied to the same initial state, while the white spot in the upper right indicates that both pulses prepare the same final state after the full pulse duration.
The bright band roughly along the diagonal of the majority of the plot indicates that both trajectories largely overlapped (except in the beginning), corroborating the qualitative picture in Fig.~\ref{fig::pathlengths:cartoon}.


\subsection{Polar vs Cartesian Parameterization}
\label{sec::polarcartesian}

\begin{figure}
    \subfloat[\label{fig::polarcartesian:energy}]{
        \includegraphics[width=\columnwidth]{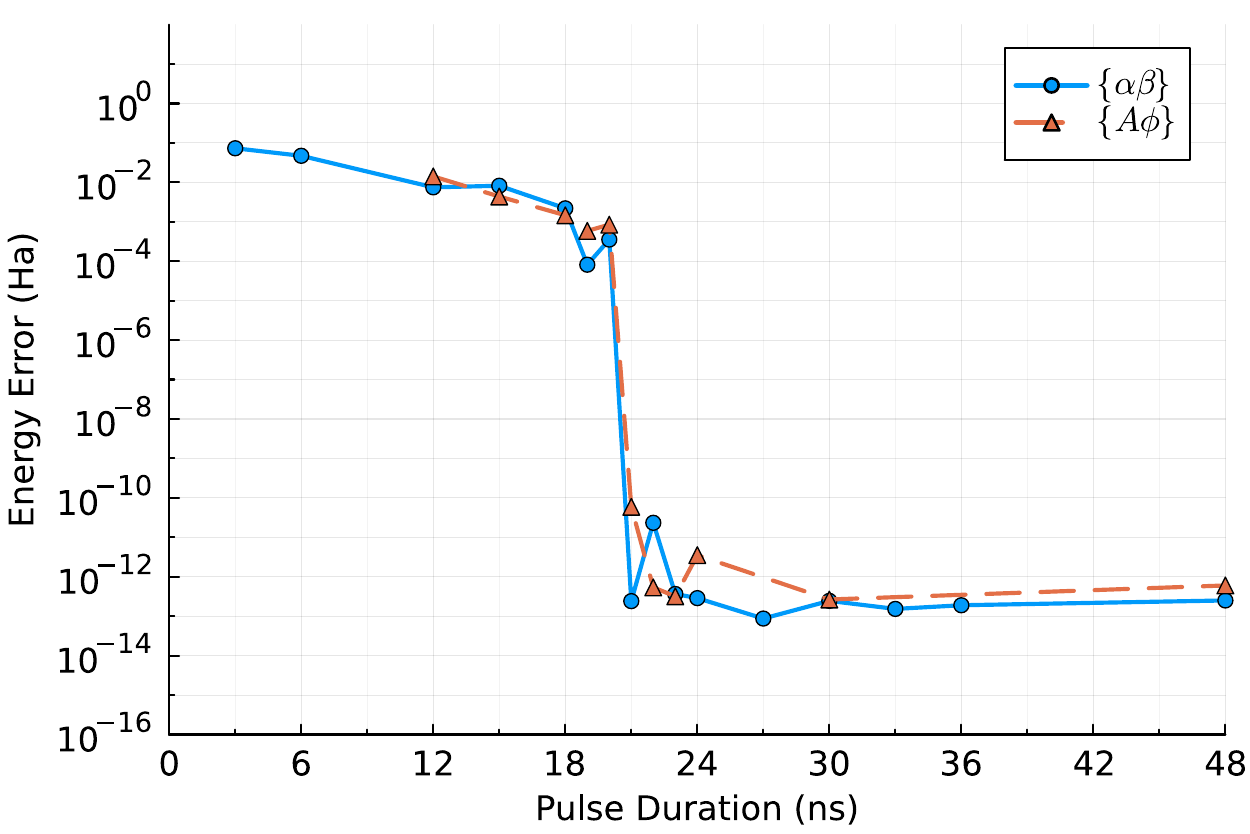}
    }
    \hfill
    \subfloat[\label{fig::polarcartesian:iterations}]{
        \includegraphics[width=\columnwidth]{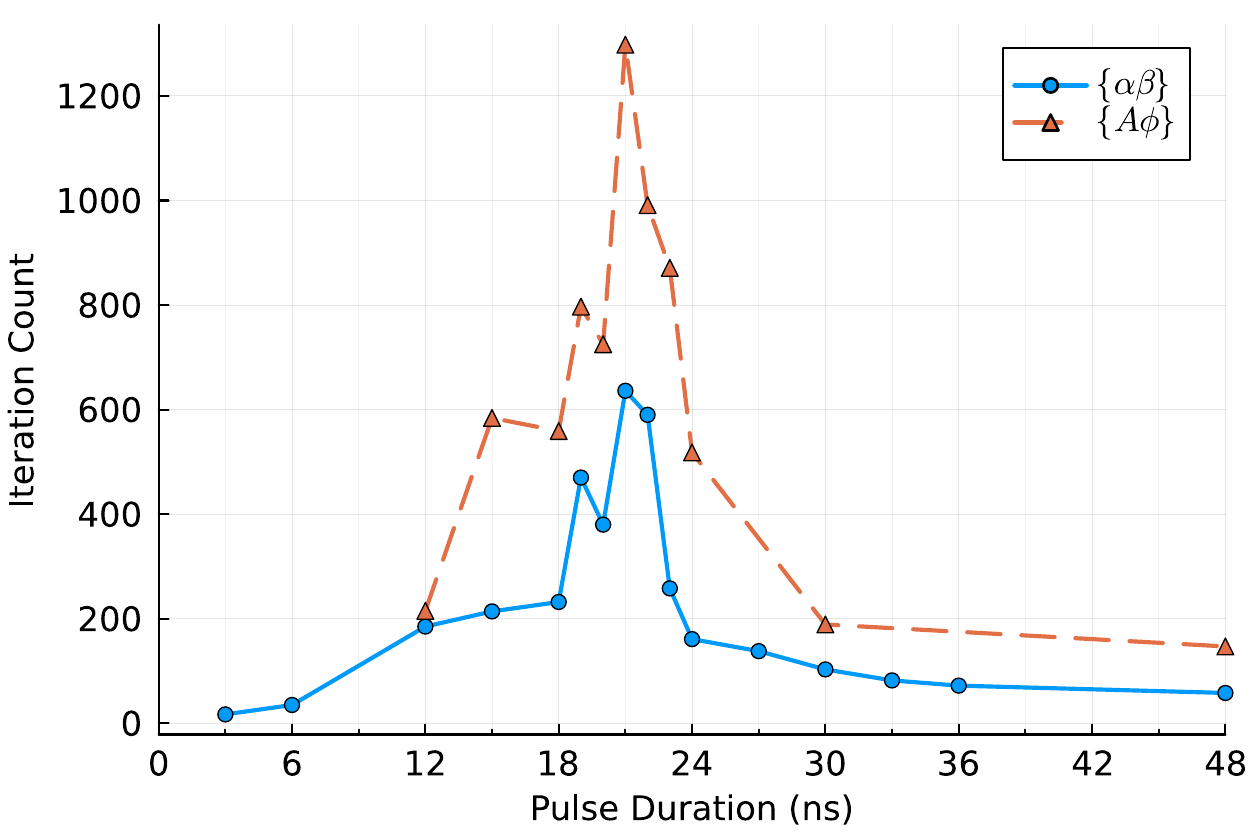}
    }
    \caption{
        (\ref{fig::polarcartesian:energy}) Ground state energy error vs. pulse duration for ctrl-VQE applied to LiH with nuclear separation 3.0\AA, parameterizing each pulse's complex amplitude with polar $\qty{A\phi}$ or Cartesian $\qty{\alpha\beta}$ notations. Each pulse is fixed on resonance and divided into as many uniformly spaced windows as possible such that the window length $\Ds\ge3.0\ns$. All parameters are initialized to zero in the optimization.
        (\ref{fig::polarcartesian:iterations}) The number of BFGS iterations needed to obtain the energies in (\ref{fig::polarcartesian:energy}).
    }
    \label{fig::polarcartesian}
\end{figure}

In this section, we contrast ctrl-VQE using Cartesian notation (Eq.~\ref{eq::cartesian}) and polar notation (Eq.~\ref{eq::polar}).
Either choice leads to the exact same dynamics for any single pulse,
    but a finite step in the different parameter sets will have different resulting pulses,
    meaning a gradient-based optimizer like BFGS
    may well proceed along different optimization trajectories.
In Fig.~\ref{fig::polarcartesian},
    we show \EvT{} and \CvT{} plots comparing the two different notations
    applied to LiH with a bond separation of 3.0\AA,
    holding frequencies fixed on resonance.
The energy plots and the eMET $\eMET$ are essentially identical,
    indicating that both runs identified pulses
    preparing essentially the same estimate of the ground state for each pulse duration $T$.
Note that the actual pulse shapes do vary, but more than one pulse generates the same state.
However, the optimization trajectories for the polar notation are notably more expensive,
    so we recommend adopting Cartesian notation.
We assume this arises due to larger off-diagonal matrix elements in the pulse Hessian of the polar representation, but this is a topic we will explore in future work.

\subsection{Importance of Phase and Frequency}
\label{sec::resonantdetuned}

\begin{figure}
    \subfloat[\label{fig::resonantdetuned:energy}]{
        \includegraphics[width=\columnwidth]{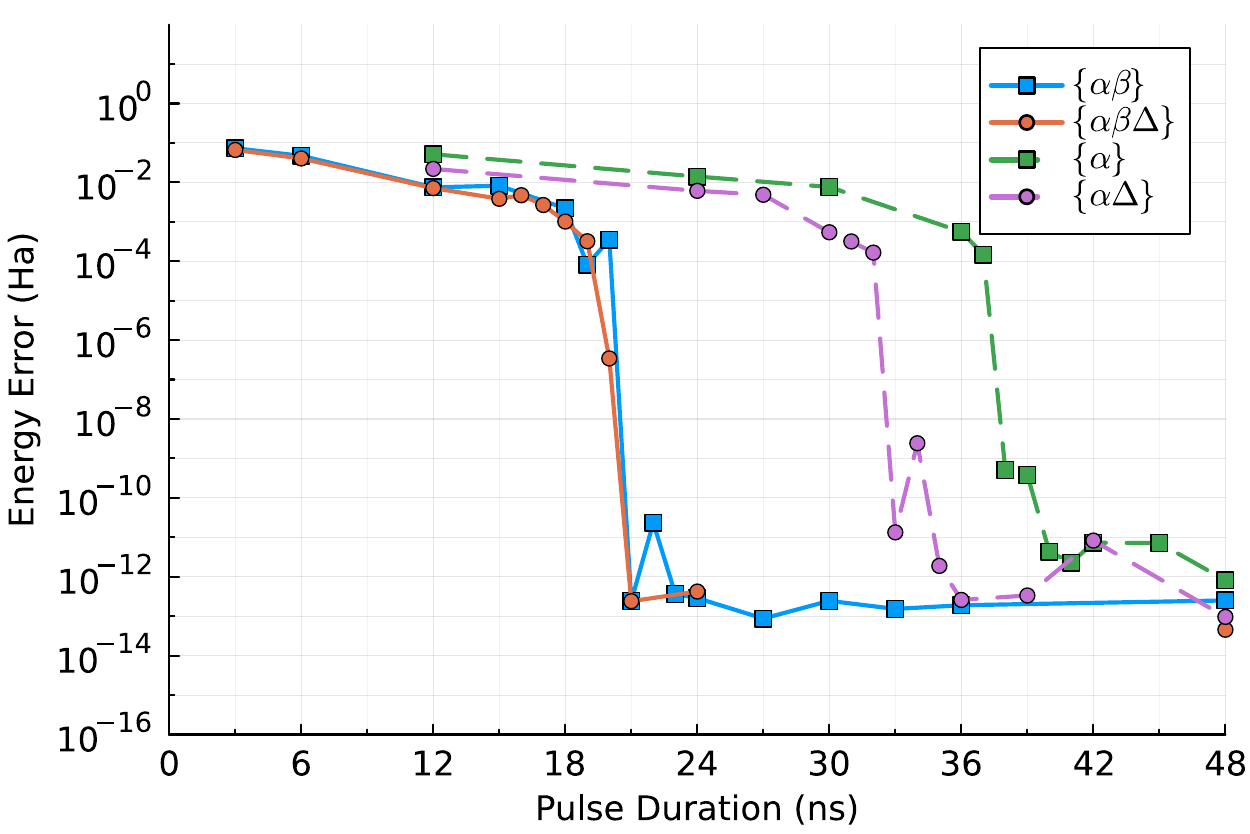}
    }
    \hfill
    \subfloat[\label{fig::resonantdetuned:iterations}]{
        \includegraphics[width=\columnwidth]{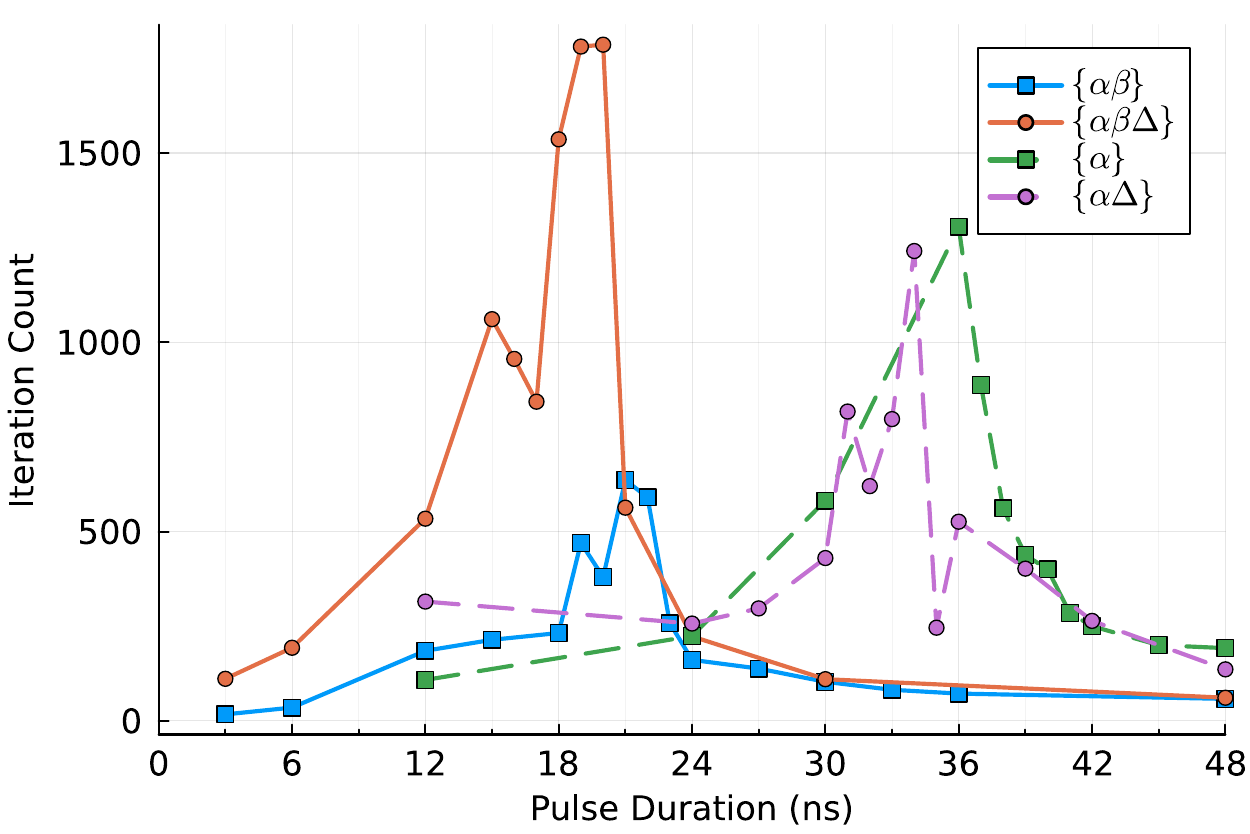}
    }
    \caption{
        (\ref{fig::resonantdetuned:energy}) Ground state energy error vs. pulse duration for ctrl-VQE applied to LiH with nuclear separation 3.0\AA, for various parameterizations. Each pulse for the $\qty{\alpha\beta}$ and $\qty{\alpha\beta\Delta}$ curves is divided into as many uniformly spaced windows as possible such that the window length $\Ds\ge3.0\ns$, while those for $\qty{\alpha}$ and $\qty{\alpha\Delta}$ use $\Ds\ge1.5\ns$, so that the total number of parameters scale similarly. All parameters are initialized to zero in the optimization.
        (\ref{fig::resonantdetuned:iterations}) The number of BFGS iterations needed to obtain the energies in (\ref{fig::resonantdetuned:energy}).
    }
    \label{fig::resonantdetuned}
\end{figure}

Previous works on ctrl-VQE \cite{Meitei2020,asthana_leakage_2023} used the $\qty{\alpha\Delta}$ parameterization,
    restricting drive phases to $\phi\in\qty{0,\pi}$ and allowing drive frequencies to vary off resonance.
In this section, we will explore what happens
    when the phase degree of freedom is included
    ($\qty{\alpha\beta\Delta}$, in Cartesian notation),
    when the frequency degree of freedom is removed ($\qty{\alpha})$,
    and when both changes are effected ($\qty{\alpha\beta}$).
Fig.~\ref{fig::resonantdetuned} gives \EvT{} and \CvT{} plots for all four parameterizations,
    applied to LiH with a bond separation of 3.0\AA.
For the sake of fair comparison,
    pulses with real-valued amplitudes are divided into twice as many windows,
    so that the total number of parameters is similar for each value of $T$.
For example, the $\qty{\alpha\beta}$ pulse at $24\ns$ is divided into 8 equal-sized windows, each with an independent $\alpha$ and $\beta$, for a total of 16 parameters. Meanwhile, the $\qty{\alpha}$ pulse at $24\ns$ is divided into 16 equal-sized windows, each with a single parameter. Parameterizations varying the frequency ($\qty{\alpha\beta\Delta},\qty{\alpha\Delta}$) include one more parameter for each pulse, so that their $24\ns$ pulses have a total of 20 parameters.

Unsurprisingly, the longest eMET $\eMET$ is observed
    with the most restrictive parameterization $\qty{\alpha}$,
    and the second longest is the second-most restrictive, $\qty{\alpha\Delta}$.
(Recall that varying $\Delta$ is more restrictive than varying $\beta$,
    since each pulse has only one $\Delta$ parameter, irrespective of the number of windows.)
Perhaps surprisingly, the remaining two parameterizations $\qty{\alpha\beta}$
    and $\qty{\alpha\beta\Delta}$ are practically indistinguishable,
    except that including the frequency degree of freedom puts a heavy strain on the optimizer (this effect was also observed in Ref.~\cite{kottmann_evaluating_2023}).
Apparently, once the pulse amplitude is allowed to become complex, there is no further variational flexibility afforded by the frequency detuning. 

We must emphasize that this is {\em not} related to the dimensionality
    of the optimization problem.
The number of degrees of freedom is similar (within a constant) for all four parameterizations.
We verify this numerically by calculating the effective quantum dimension
    \cite{haug_capacity_2021}
    of our trial state at several steps of the optimization.
This quantity is defined as the rank of the quantum Fisher information~\cite{stokes_quantum_2020},
    and it describes the number of directions in which infinitesimal variations in our parameters can drive the system.
When the pulse has a sufficient number of parameters, we anticipate the effective quantum dimension to saturate at the number of independent degrees of freedom in the Hilbert space (30, for a four-qubit system).
Computational basis states are {\em singular}, isolated points with reduced dimension.
Thus, we expect (and find) initial pulses, with $\Omega=0$, to have an effective quantum dimension of 16,
    irrespective of the number of parameters in the pulse.
However, random perturbations from a singular point
    are almost sure to result in a state with full effective quantum dimension~\cite{haug_capacity_2021,funcke_dimensional_2021},
    and indeed we find that the effective quantum dimension is saturated at 30 for all subsequent trial states when $T\ge12\ns$.

It is natural to wonder why the frequency degree of freedom has no apparent impact on the eMET,
    so long as we include the phase degree of freedom
    (i.e., why the $\qty{\alpha\beta}$ and $\qty{\alpha\beta\Delta}$ parameterizations are so similar).
This can be {\em partly} explained by noting that the drive phase $\phi$
    and drive frequency $\nu$ are intrinsically related:
    the argument to the trigonometric function in Eq.~\eqref{eqn:EM} is the sum $\nu t+\phi$.
In the limit where the phase $\phi$ is allowed to vary continuously,
    one could implement perturbations to the drive frequency $\nu\rightarrow\nu+\Delta$
    without actually changing the drive frequency directly,
    by selecting $\phi \rightarrow \phi + \Delta t$ \cite{PATT199294}.
Consequently, we observe essentially identical eMETs with and without a frequency degree of freedom
    even for finite window lengths, where the phase $\phi$ cannot have a linear dependence on time.
Section~\ref{sec::bloch} analytically explores a toy problem to address this question more thoroughly.

\subsubsection{Phase versus Frequency for a Single Qubit}
\label{sec::bloch}

In this section, we will consider the analytical solution
    for a single qubit driven by a pulse with a constant amplitude $A$, detuning $\Delta$, and phase $\phi$.
Our objective is to obtain a qualitative understanding of why the phase $\phi$ alone is sufficient to minimize the eMET, as seen in Fig.~\ref{fig::resonantdetuned}.
We will describe our single-qubit quantum state using the Bloch sphere parameterization $\qty(\theta,\varphi)$:
\begin{equation}
    \ket{\psi} = \cos{\frac{\theta}{2}} \ket{0} + e^{i\varphi}\sin{\frac{\theta}{2}} \ket{1}.
\end{equation}
The objective in ctrl-VQE is to prepare a particular state from a reference state;
    in this example, let us take the north pole $\qty(0,0)$ as our reference state ($\ket{\psi_\REF}=\ket{0}$),
    and the target state corresponds to the point $\qty(\theta_0,\varphi_0)$. 

\begin{figure*}
    \subfloat[\label{fig::cartoon:PD:00}]{
        \includegraphics[width=0.3\textwidth]{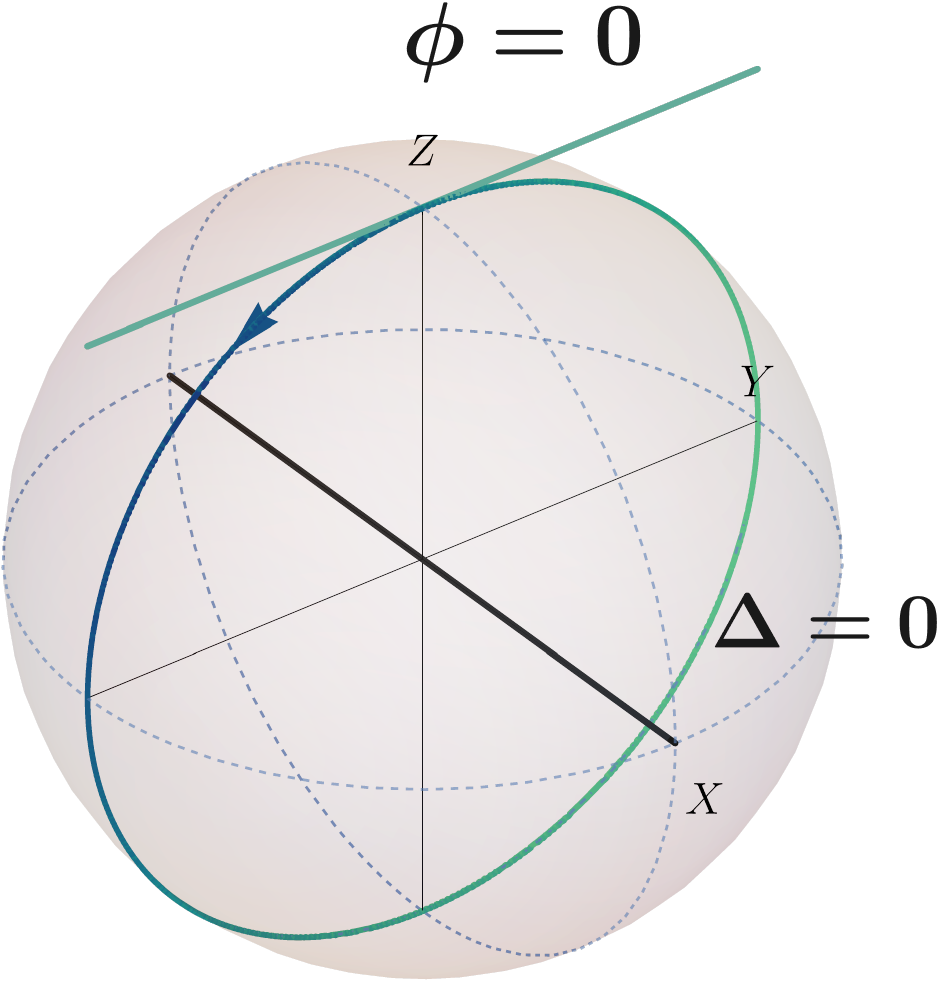}
    }
    \hfill
    \subfloat[\label{fig::cartoon:PD:10}]{
        \includegraphics[width=0.3\textwidth]{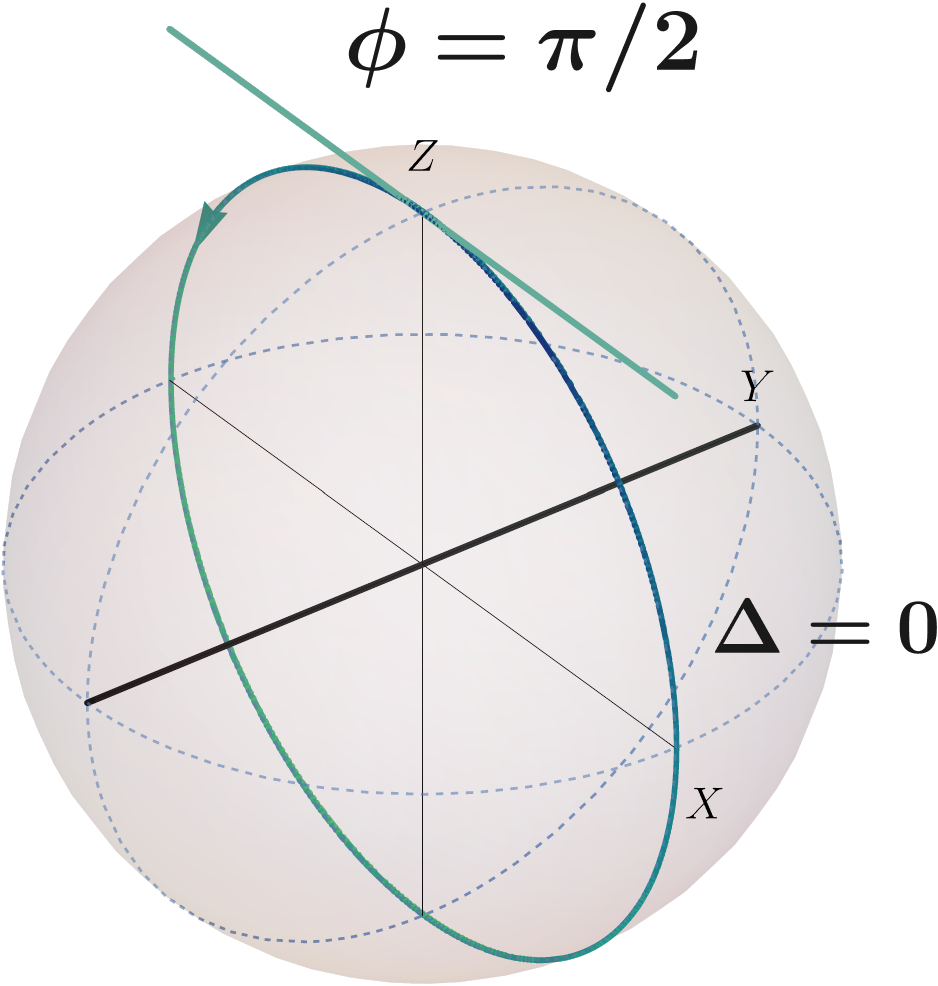}
    }
    \hfill
    \subfloat[\label{fig::cartoon:PD:11}]{
        \includegraphics[width=0.3\textwidth]{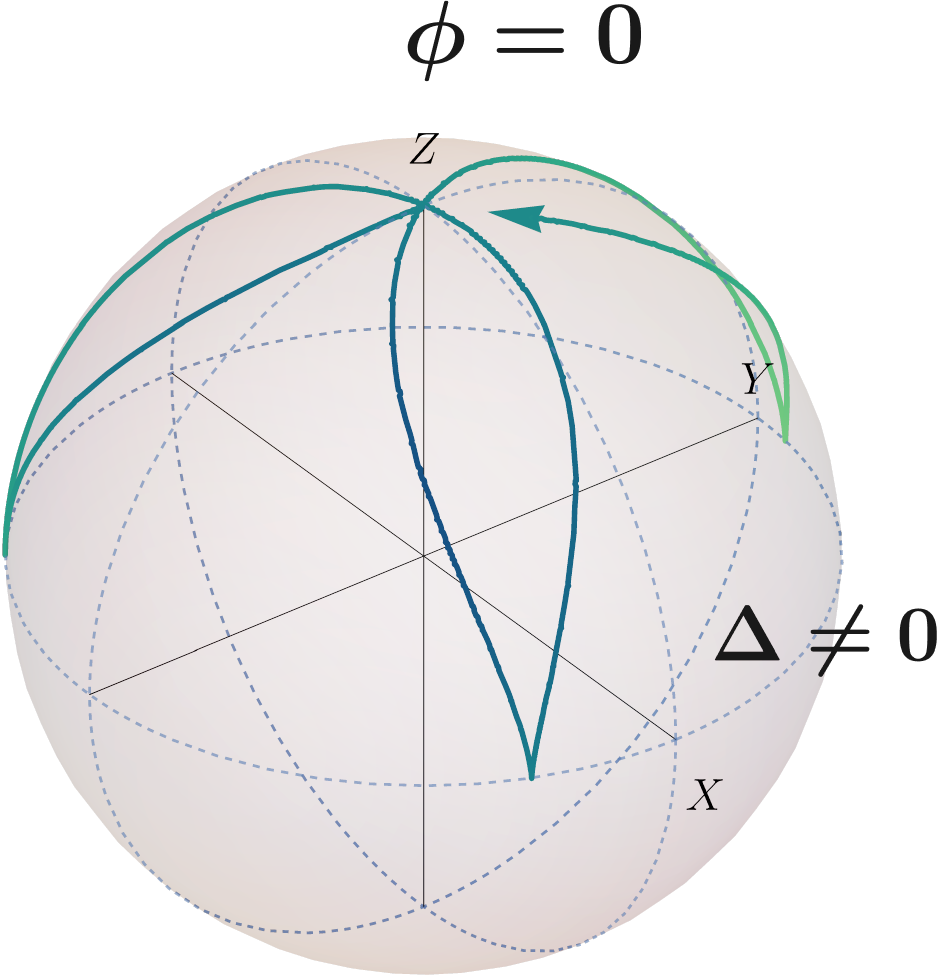}
    }

    \caption{Cartoon depictions of the path on a Bloch sphere traced by a single-qubit state in the rotating frame, under a pulse with constant $A$, $\phi$, and $\Delta$, when starting from the north pole ($\ket{0}$). The path rotates around an axis with azimuthal orientation starting at $\phi$ and precessing at a rate $\Delta$, and a vertical tilt given by the ratio $A/\Delta$. When $\Delta=0$ (\ref{fig::cartoon:PD:00} and \ref{fig::cartoon:PD:10}), the axis is in the $x$-$y$ plane, so the path is a meridian with longitude given by $\phi$.}
    \label{fig::cartoon:PD}
\end{figure*}

In the rotating frame, and under the rotating wave approximation,
    the system evolves under the Hamiltonian in Eq.~\eqref{eqn:rotatingdrive}
    according to the  time-dependent Schr\"odinger equation, which can be solved exactly.
This is the Rabi problem, which can be found in standard quantum mechanics textbooks, e.g. see Ref. \cite{townsend2000modern}.
The trajectories of the qubit state on the Bloch sphere in the rotating frame are visualized for several choices of $\phi$ and $\Delta$ in Fig.~\ref{fig::cartoon:PD}.
The state rotates around an axis
    oriented along a polar angle $\arctan(A/\Delta)$
    and an azimuthal angle that begins along $\phi$ but precesses slowly in time at a rate $\Delta$.
In addition to informing the axis of rotation, the amplitude $A$ also determines the speed at which this path is traversed,
    so we will always want it as large as practically obtainable
    in order to minimize the necessary evolution time.

When the detuning vanishes, $\Delta=0$, the axis of rotation is in the $x$-$y$ plane;
    since we start from the north pole, the trajectory is a meridian,
    with the longitude fixed by the pulse phase $\phi$.
Therefore, the azimuthal coordinate of our state $\varphi$ is constant (up to $\pm\pi$).
The most direct state preparation strategy
    is to choose $\phi=\varphi_0$ and drive on resonance ($\Delta=0$).
Since the shortest path (i.e., the geodesic)
    between a pole and any point on a sphere is along the meridian,
    this choice guarantees the shortest possible evolution time,
    subject to constraints on the amplitude $A$.

If we restrict ourselves to real-valued pulses as in the previous section
    (i.e., $\phi\in\qty{0,\pi}$),
    the only azimuthal coordinates we can access with resonant pulses are
    $\varphi\in\qty{0,\pi}$ (see Fig.~\ref{fig::cartoon:PD:00}).
If our target state has any other azimuthal coordinate $\varphi_0$,
    we {\em must} use an off-resonant pulse (i.e., $\Delta \ne 0$).
Doing so tilts the angle of the axis about which our trajectory rotates;
    the trajectory is no longer a geodesic.
Thus, not only are large polar angles $\theta>2\arctan(A/\Delta)$ entirely inaccessible,
    but target states with smaller $\theta$ will require a longer path
    and therefore a longer pulse duration.

To summarize, reaching any arbitrary single-qubit state from the $\ket{0}$ state requires varying either the drive's phase $\phi$ or frequency $\Delta$.
Varying $\phi$ allows us to select an initial direction to move in Hilbert space,
    while varying $\Delta$ allows us to introduce a curve in the trajectory.
In multi-qubit systems, the ideal trajectory is harder to characterize,
    since the optimal trajectory using the control fields accessible to us experimentally do not necessarily trace a geodesic in the full Hilbert space.
However, the freedom to adjust phases will tend to allow us to select a more direct path, while adjusting frequency will tend to incur a longer path length which takes longer to traverse.

\section{Conclusions}
\label{sec:conclusions}

In this paper, we have explored the consequences of several different methods
    to parameterize a pulse-level ansatz for VQE experiments.
For each parameterization, we analyzed how energy accuracy and optimization difficulty vary over pulse duration.
Here, we review our most salient observations and summarize our main conclusions.

In every case, it is clear that ctrl-VQE requires a minimal pulse duration
    to have any chance of successfully preparing the desired eigenstate.
While operating at this minimal time would be ideal for minimizing qubit decoherence,
    and it would seem to minimize runtime,
    we find that the optimization loop takes much longer to converge at and near this regime.
Furthermore, even locating the MET (or eMET) would be intractable in practice. 
Consequently, a challenge arises for realistic simulations: how does one know \textit{a priori} that the specified evolution time in fact exceeds the MET, and thus provides the possibility of preparing the target state? 
While answering this definitively would lie beyond the scope of this paper, we note that the quality of the final solution can be estimated by metrics such as variance \cite{zhang_variational_2022}
or Hamiltonian-reconstruction distance \cite{moon_hamiltonian_2024}
to determine whether the optimization has located an eigenstate.

In Section~\ref{sec::windowspacing},
    we explored the consequences of restricting pulse parameters
    from continuous functions in time to discrete windows.
So long as individual windows are not too long,
    we found that ctrl-VQE provides essentially consistent results, indicating that digitization of the analogue signals should not have a significant impact on the performance.
The minimal evolution time required to obtain high-accuracy ans\"atze
    increases marginally as the number of free parameters is reduced,
    but the far more significant difference is that
    optimization is harder when the parameterization better approximates continuously variable pulses.
In some sense, this is a trivial statement: optimization with fewer parameters is easier.
Nevertheless, it has important practical significance,
    since the runtime of ctrl-VQE is ultimately bounded by the number of optimization iterations.

Similarly, in Section~\ref{sec::polarcartesian},
    we tested parameterizations for a complex amplitude
    represented in both polar ($\Omega=Ae^{i\phi}$)
    and Cartesian ($\Omega=\alpha+i\beta$) coordinates.
These two choices are physically identical;
    the only difference between them is the trajectory through parameter space
    which the optimizer selects to find optimal pulses.
This is analogous to the situation in classical molecular geometry optimization, where different coordinate representations ($z$-matrix, redundant internals, Cartesians, etc.) all specify the same molecule, but lead to widely varying iteration counts for optimization.  \cite{baker1996}
We find that the polar parameterization turns out to be significantly more expensive to converge,
    especially in the region near the minimal evolution time.
Therefore, we recommend using Cartesian coordinates for experimental implementations of ctrl-VQE.

Our most important results are found in Section~\ref{sec::resonantdetuned},
    where we contrast parameterizations which allow or disallow
    the phase and the frequency to vary in the optimization.
We find that the minimal evolution time is longest when only amplitudes are varied,
    that it is improved when amplitudes and frequencies are varied,
    and that it is improved even more when amplitudes and phases are varied.
Counter-intuitively, we find {\em no} measurable improvement to the minimal evolution time
    when all three parameters are varied,
    although the optimization is disproportionately harder.
These results suggest that, in a transmon-based ctrl-VQE, it is preferable to fix the drive frequencies of each pulse (perhaps  at the associated resonance frequency) and optimize over amplitudes and phases.

Since single-qubit gates are often implemented with resonant pulses,
    it may be surprising that resonant pulses are capable of building the entanglement needed to represent the ground state of a molecular system.
However, in an architecture with static couplings,
    pulses applied locally in the ``qubit'' basis will nevertheless
    have a global effect in the eigenbasis of the device.
In some sense, entanglement builds spontaneously, irrespective of pulse frequency;
    the pulses serve only to direct it in the appropriate direction
    and drive it along faster.
\textit{Our findings imply that the phase of control fields is the more effective parameter to provide that direction.}
Modulating this parameter in time is also more natural in fixed-frequency, fixed-coupling devices, such as those presently used by IBMQ~\cite{mckay_efficient_2017}.
An interesting direction for future research is to investigate
    whether similar or complementary heuristics can be found for alternate architectures,
    such as those with tunable couplers, or those where the drive field is applied globally to all qubits.

Another pressing research question for pulse-level VQE is its scalability to larger systems - namely, whether resonant pulses maintain the orders-of-magnitude improvement over gate-based VQE as the number of qubits increases, and if there is a consistent strategy to further reduce the number of parameters needed to obtain accurate ans\"atze. These important questions are being addressed in ongoing and future work.


\begin{acknowledgements}

The authors thank Chenxu Liu, Ayush Asthana, and Christopher Long for edifying conversations.
This research was supported by the U.S. Department of Energy (DOE) Award No. \textbf{DE-SC0019199}, by the National Science Foundation (Grant No. 2137776). 
S.E.E. acknowledges support from the DOE Office of Science, National Quantum Information Science Research Centers, Co-design Center for Quantum Advantage (C2QA)
    under contract number \textbf{DE-SC0012704}.

\end{acknowledgements}


\section{Supplementary Information}

Our latest implementation of ctrl-VQE can be found at \cite{CtrlVQE}.



\bibliography{complexctrl}

\appendix


\section{Evaluating Analytical Gradients with Complex Amplitudes}
\label{app::gradient}

In this appendix, we derive an expression for the analytic gradient
    of the expectation value $\expval{\hat O}{\psi}$
    with respect to our control parameters $\x$,
    where $\ket{\psi}=\ket{\psi(\x)}$ is our fully-evolved state
    and $\hat O$ is a Hermitian observable.
This treatment is an adaptation of GRAPE \cite{khaneja_optimal_2005},
    tailored to our particular problem.

Formally, the objective function takes the form:
\begin{equation}
    E(\x) = \expval{\hat O}{\psi_R(\x)}
\end{equation}
By using a subscript $R$ in $\ket{\psi_R(\x)}$,
    we denote that $\hat O$ is measured in the rotating frame.

The variational ansatz takes the form:
\begin{equation}
    \ket{\psi_R(\x)} = U_R(\x) \ket{\psi_\REF}
\end{equation}
where $U_R(\x)$ is the unitary operator describing evolution under the device Hamiltonian:
\begin{equation}
    U_R(\x) = \mathcal{\hat T} \exp(-i \int_0^T \dif{t} \hat V_R(\x, t))
\end{equation}
The device Hamiltonian itself (in the rotating frame) takes the following form:
\begin{equation}
    \hat V_R(\x, t) = e^{i t \hat H_0} \hat V(\x,t) e^{-i t \hat H_0}
\end{equation}
The static component $\hat H_0$ does not depend on our control parameters,
    so it need not be specified further in this derivation.
The drive component $\hat V$ has the form:
\begin{equation}
    \hat V(\x,t) = \sum_q \hat V_q(\x,t) = \sum_q \qty( \Omega_q(\x, t) e^{i\nu_q(\x,t)t} \an_q + \ha )
\end{equation}
Recall that $\Omega_q$ is a complex number,
    encapsulating both the amplitude and phase of the drive pulse.
It will be helpful to recast the drive component into a sum
    of Hermitian time-independent operators with real scalar time-dependent coefficients:
\begin{equation}
    \hat V(\x,t) = \sum_q \qty( x_q(\x,t) \hat Q_q + y_q(\x,t) \hat P_q )
\end{equation}
Here $\hat Q_q \equiv \an_q + \at_q$ and $\hat P_q \equiv i(\an_q - \at_q)$
    are the canonical coordinate and momentum operators,
    equivalent to Pauli-$X$ and $Y$ spin operators in the two-level truncation
    (except that our definition of $\hat P$ absorbs an extra negative sign
    for the sake of notational convenience).
The scalar functions $x_q$ and $y_q$ are the real and imaginary parts of $\Omega_q \exp(i\nu_q t)$.

The simplest approach to characterizing time-dependent functions is to discretize time.
Therefore, let us define a time-interval $\tau\equiv T/r$, where $r$ should be understood as arbitrarily large.
We will abbreviate all parametric time-dependent $f(\x,t)$ with the shorthand:
\begin{equation}
    f_i \equiv f(\x,i\tau)
\end{equation}
Now we choose to rewrite the integral in the exponential of $U_R$ as a trapezoidal sum.
\begin{equation}
    \label{eq:gradient:traprule}
    \int_0^T \dif{t} \hat V_R(\x,t) \approx \tau \sum_{j=1}^r \frac{\hat V_{R,j} + \hat V_{R,j-1}}{2}
\end{equation}
Furthermore, since $V_R$ is a conjugation of the drive component $\hat V$
    with respect to a unitary frame rotation $\exp(-it\hat H_0)$,
    this factor can be brought {\em outside} the exponentiation $\exp(-it\hat V_R)$:
\begin{equation}
    e^{-i\tau \hat V_{R,i}} = e^{it\hat H_0} e^{-i\tau \hat V_i} e^{-it\hat H_0}
\end{equation}
We may now expand $U_R$ as a time-ordered product.
At the same time, we will introduce left and right unitaries $U_{Ri}^{(l)}$, $U_{Ri}^{(r)}$
    symmetrically about a single time point $t=i\tau$:
\begin{align}
    U_R &\equiv U_{Ri}^{(l)} U_{Ri}^{(r)} \text{~~~~~for any $i\in0..r$}\\
    U_{Ri}^{(l)} &\approx e^{iT \hat H_0} \oprod_{j=i+1}^{r}
        \qty( e^{-i\tau\hat V_j / 2} e^{-i\tau\hat H_0} e^{-i\tau\hat V_{j-1} / 2} ) \\
    U_{Ri}^{(r)} &\approx \oprod_{j=1}^{i}
        \qty( e^{-i\tau\hat V_j / 2} e^{-i\tau\hat H_0} e^{-i\tau\hat V_{j-1} / 2} )
\end{align}

In order to take the gradient with respect to control parameters $\x$,
    let us first consider partial derivatives with respect to $x_{qi}$ and $y_{qi}$.
We will begin by writing $\hat V_i= \hat\Sigma + x_{qi} \hat Q_q$,
    where $\hat \Sigma$ represents all the terms in $\hat V_i$ except $x_{qi} \hat Q_q$.
The analytical expression for the derivative of the exponential of $\hat V_i$ is \cite{khaneja_optimal_2005}:
\begin{equation}
    \partial x_{qi} e^{-i\tau \qty(\hat\Sigma + x_{qi} \hat Q_q)}
        = e^{-i\tau \hat\Sigma} \int_0^1 \dif{s}
            e^{-i\tau s \hat\Sigma}
            \qty(-i\tau \hat Q_q)
            e^{i\tau s \hat\Sigma}
\end{equation}
Because $\hat Q_q$ and $\hat P_q$ do not commute
    (in the two-level truncation, the commutator is not even scalar),
    this has no closed-form solution.
However, so long as $\tau$ is very small, corrections due to the commutator are negligible.
Therefore, we choose to adopt the following much simpler, approximate expressions:
\begin{align}
    \partial_{x_{qi}} U_{Ri}^{(l)} &\approx U_{Ri}^{(l)} \qty(-i\tau \hat Q_q/2) \\
    \partial_{x_{qi}} U_{Ri}^{(r)} &\approx \qty(-i\tau \hat Q_q/2) U_{Ri}^{(r)} \\
    \partial_{x_{qi}} U_R &\approx -i\tau U_{Ri}^{(l)} \hat Q_q U_{Ri}^{(r)}
\end{align}
Similarly,
\begin{equation}
    \partial_{y_{qi}} U_R \approx -i\tau U_{Ri}^{(l)} \hat P_q U_{Ri}^{(r)}
\end{equation}
Note that the symmetric form of our trapezoidal time integration
    serves to further enhance the accuracy of the approximation.

We now define the ``gradient signals'' $\phi_{qi}^{(x)}$, $\phi_{qi}^{(y)}$:
\begin{align}
    \phi_{qi}^{(x)} &= \Im \expval{\adj{U_R} \hat O U_{Ri}^{(l)} \hat Q_q U_{Ri}^{(r)}}{\psi_\REF} \\
    \phi_{qi}^{(y)} &= \Im \expval{\adj{U_R} \hat O U_{Ri}^{(l)} \hat P_q U_{Ri}^{(r)}}{\psi_\REF}
\end{align}
These can be computed numerically
    in a time comparable to that of simulating time-evolution itself.
It is straightforward to show that the gradient of the objective function $E$
    with respect to $x$ and $y$ is:
\begin{align}
    \partial_{x_{qi}} E &= 2\tau \phi_{qi}^{(x)} \\
    \partial_{y_{qi}} E &= 2\tau \phi_{qi}^{(y)}
\end{align}
Finally, a simple chain rule gives us the expression for the energy gradient
    with respect to our control parameters $\x$:
\begin{equation}
    \partial_{\theta_k} E = 2\tau \sum_{q,i} \qty[
        \qty(\partial_{\theta_k} x_{qi}) \cdot \phi_{qi}^{(x)}  
        + \qty(\partial_{\theta_k} y_{qi}) \cdot \phi_{qi}^{(y)}
    ]
\end{equation}
We note that the above formula is in effect a discretized integral over time,
    and accuracy can be tuned by adopting any quadrature rule
    (e.g., the trapezoidal rule used in Eq.~\eqref{eq:gradient:traprule}).
    We conclude by explicitly writing the integral formulation for continuous time:
\begin{align}
    \partial_{\theta_k} E = 2 \sum_q \int_0^T \dif{t}& 
    \left[
        \fullpartial{x_q(\x,t)}{\theta_k} \cdot \phi_{q}^{(x)}(t)\right.\nonumber\\        
        &\left.+ \fullpartial{y_q(\x,t)}{\theta_k} \cdot \phi_{q}^{(y)}(t)\right]
\end{align}

\section{ctrl-VQE with Larger Systems}
\label{sec::longersurvey}

\begin{figure}
    \subfloat[\label{fig::longersurvey:energy}]{
        \includegraphics[width=\columnwidth]{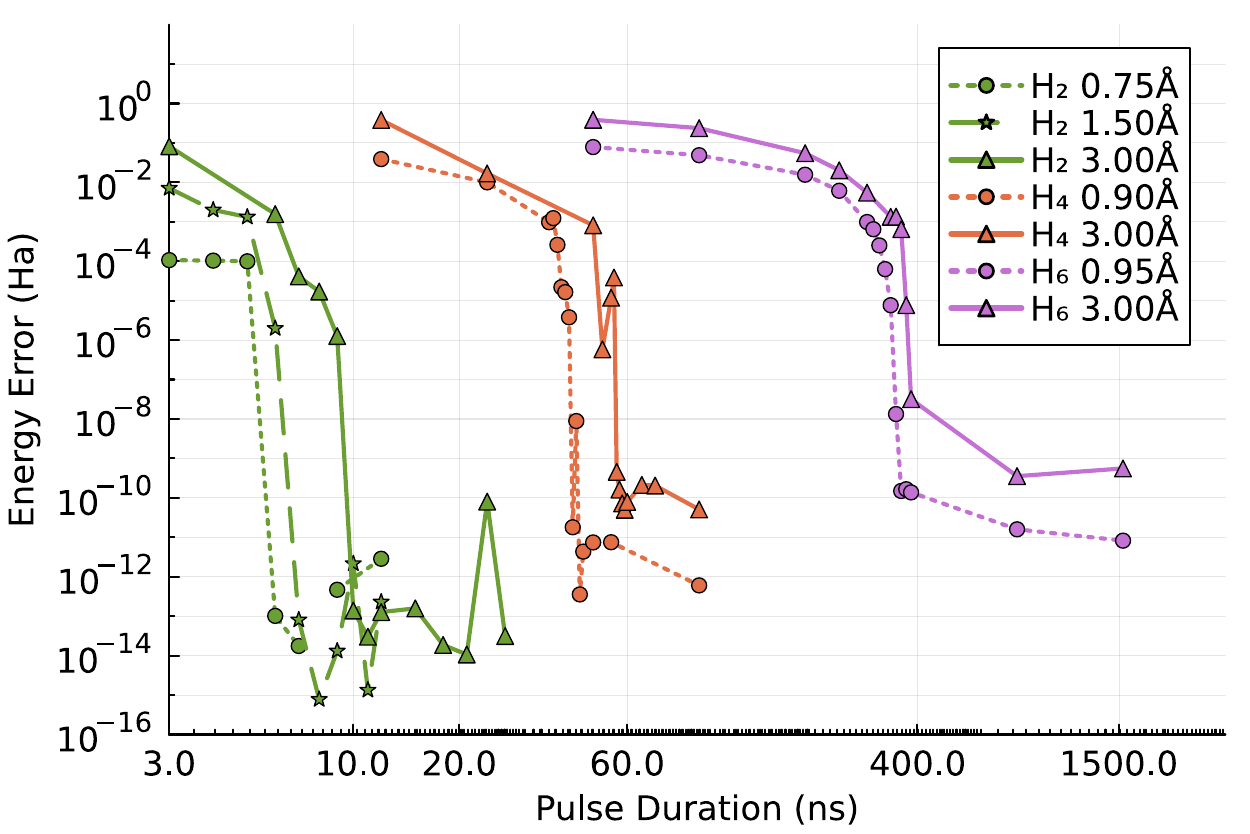}
    }
    \hfill
    \subfloat[\label{fig::longersurvey:iterations}]{
        \includegraphics[width=\columnwidth]{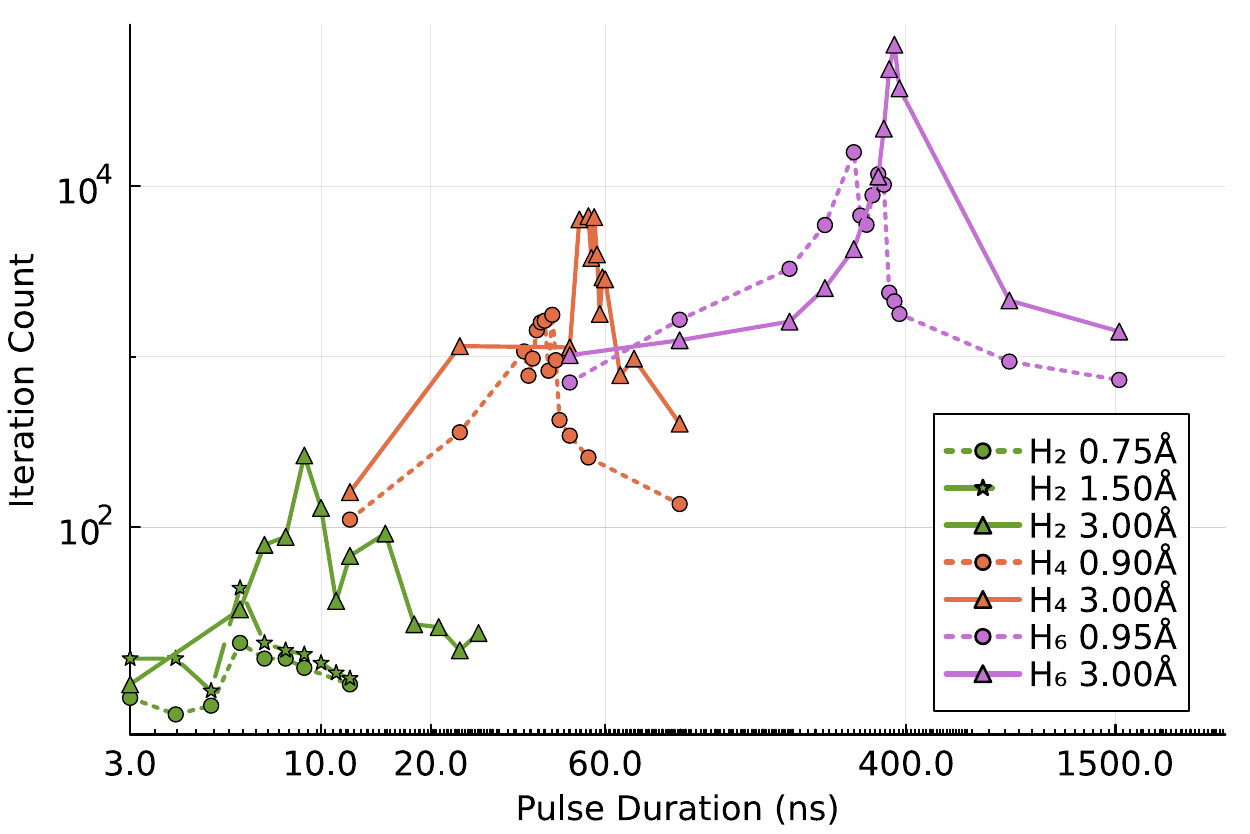}
    }
    \caption{
        (\ref{fig::longersurvey:energy}) Ground state energy error vs. pulse duration of ctrl-VQE applied to increasingly larger molecules with bond distance specified in the legend, using the $\qty{\alpha\beta}$ parameterization. Each pulse is fixed on resonance and divided into as many uniformly spaced windows as possible such that the window length $\Ds\ge3.0\ns$. All parameters are initialized to zero in the optimization.
        (\ref{fig::longersurvey:iterations}) The number of BFGS iterations needed to obtain the energies in (\ref{fig::longersurvey:energy}).
    }
    \label{fig::longersurvey}
\end{figure}

In Fig.~\ref{fig::longersurvey}, we show preliminary results from applying the lessons we have learned in this paper to larger systems - $H_4$ mapped to six qubits, and $H_6$ mapped to ten qubits.
We also show results for $H_2$ mapped to two qubits.
In Fig.~\ref{fig::longersurvey:energy}, we plot the x-axis on a log-scale.
We see that the effective MET $\eMET$ is somewhat influenced by the molecular geometry, but most especially by the system size.
Indeed, the increase in $\eMET$ appears to be roughly exponential in the number of qubits.
Nevertheless, the actual pulse durations necessary to prepare the $H_6$ ground state with these device parameters may be as much as 2000 times faster than pulse compilation of the most sophisticated gate-based approaches \cite{tang2021qubit}.
We plot Fig.~\ref{fig::longersurvey:iterations} with the y-axis also on a log-scale, highlighting that the optimization problem near $\eMET$ appears exponential with pulse duration.
Nevertheless, operating in the regime somewhat beyond $\eMET$ should allow for classically tractable optimization while preserving the lion's share of pulse-based VQE over gate-based VQE.

\end{document}